\documentclass{elsart}
\usepackage{times}
\usepackage{cite}
\usepackage{graphicx}
\usepackage{color}
\usepackage{subfigure}
\usepackage{amssymb}
\usepackage{amsmath}
\usepackage{txfonts}
\usepackage{calrsfs}
\usepackage{ulem}

\def\mathbi#1{\textbf{\em #1}}

\begin{document}
\normalem

\begin{frontmatter}

\title{Performance of Random Walks in One-Hop Replication Networks}

\author[tid]{Luis Rodero-Merino\corauthref{cor1}},
\ead{rodero@tid.es}
\corauth[cor1]{Corresponding author. Tel: +34 913373928, fax: +34
913374272, postal address: Telef\'{o}nica I+D, C/Emilio Vargas 6, C.P.:28043,
Madrid, Spain.}
\author[urjc]{Antonio Fern\'{a}ndez Anta},
\ead{anto@gsyc.es}
\author[urjc]{Luis L\'{o}pez},
\ead{llopez@gsyc.es}
\author[uji]{Vicent Cholvi}
\ead{vcholvi@uji.es}
\address[tid]{Telef\'{o}nica I+D, Madrid, Spain}
\address[urjc]{LADyR, GSyC, Universidad Rey Juan Carlos, M\'{o}stoles, Spain}
\address[uji]{Universitat Jaume I, Castell\'{o}n, Spain}

\begin{abstract}
Random walks are gaining much attention from the networks research community.
They are the basis of many proposals aimed to solve a variety of network-related
problems such as resource location, network construction, nodes sampling, etc.
This interest on random walks is justified by their inherent properties.  They
are very simple to implement as nodes only require local information to take
routing decisions.  Also, random walks demand little processing power and
bandwidth.  Besides, they are very resilient to changes on the network topology.

Here, we quantify the effectiveness of random walks as a search mechanism in
\emph{one-hop replication networks}: networks where each node knows its
neighbors' identity/resources, and so it can reply to queries on their behalf.
Our model focuses on estimating the expected average search time of the random
walk by applying network queuing theory.  To do this, we must provide first the
expected average search length.  This is computed by means of estimations of the
expected average coverage at each step of the random walk. This model takes into
account the \emph{revisiting effect}: the fact that, as the random walk
progresses, the probability of arriving to nodes already visited increases,
which impacts on how the network coverage evolves. That is, we do not model the
coverage as a memoryless process. Furthermore, we conduct a series of
simulations to evaluate, in practice, the above mentioned metrics. Our results
show a very close correlation between the analytical and the experimental
results.
\end{abstract}


\end{frontmatter}

\section{Introduction}
\label{sec:intro}
\emph{Random walks} are a mechanism to route messages through a network. At each
hop of the random walk, the node holding the message forwards it to some
neighbor chosen uniformly at random. Random walks have interesting properties:
they produce little overhead and network nodes require only local information to
route messages. In turn, this makes random walks resilient to changes on the
network structure.  Thanks to these features, random walks are useful for
different applications, like routing, searching, sampling and self-stabilization
in diverse distributed systems such as Peer-to-Peer (P2P) and wireless
networks~\cite{Gkantsidis04b, Chawathe03, Lv02, Lv02b, Bisnik05,
Mabrouki07,Alanyali06, Dolev06, Law03, Sadagopan05}.

Past works have addressed the study of random walks.  Some of this research has
focused on the coverage problem, trying to find bounds for the expected number
of hops taken by a random walk to visit all vertices (nodes) in a
graph\footnote{The term \emph{time} to refer to the number of hops of the random
walk (that is, its \emph{length}) is usual in many previous works.  Thus, for
example, $C_G$ is often denoted the \emph{cover time}. However, in this work we
will use the term \emph{time} to refer to the \emph{duration} of the random
walk. To avoid confusion, from now on the term \emph{time} will only denote the
physical magnitude.} $G$ ($C_G$)~\cite{Feige95,Kahn88,Zuckerman90,Aldous89}.
Results vary from the optimal $C_G$ of complete graphs $\Theta(n\log
n)$~\cite{Feige95} (where $n$ is the number of vertices) to the worst case found
in the lollipop graph $\Theta(n^3)$~\cite{Motwani95}. Barnes and Feige in
\cite{Barnes93} generalize this bound to the expected number of hops to cover a
fraction ($f<n$) of the vertices of the network, which they found is
$\Theta(f^3)$. Other works, for example, are devoted to find bounds on the
expected number of steps before a given node $j$ is visited starting from node
$i$ ($H_{i,j}$). For example, it is known that the upper bound for $H_{i,j}$ is
$\Theta(n^3)$~\cite{Brightwell89}.  Many of these results are based on the study
of the properties of the transition matrix $\mathbi{P}$ and adjacency matrix
$\mathbi{A}$ in spectral form~\cite{Lovasz93}.

The previous results are used in several works to discuss the properties of
random walks in communication networks. Gkantsidis et al.~\cite{Gkantsidis04}
apply them to argue that random walks can simulate random sampling on P2P
networks, a property that in their opinion justifies the `success of the random
walk method' when proposed as a search tool~\cite{Lv02} or as a network
constructing method~\cite{Law03}.  Adamic et al.~\cite{Adamic01} study the
search process by random walks in power-law networks applying the generating
function formalism. This work seems deeply inspired by a previous contribution
of Newman et al.~\cite{Newman01}, who study the properties (mean component size,
giant component size, etc.) of random graphs with arbitrary degree distribution. 

This paper introduces a study of random walks from a different perspective. It
does not study the formal bounds in the amount of hops to cover the network.
Instead, it tries to estimate the efficiency of the random walk as a search
mechanism in communications networks, applying network queuing theory. It takes
into account the bounded processing capacities of the nodes of the network and
the load introduced by the search messages, that are routed using random walks.
To obtain this load, we need to estimate first the average search length, which
in turn is computed from the expected average coverage: the average number of
different nodes covered at each hop of the random walk. A distinguishing feature
of our work is that, as in the case of Adamic et al.~\cite{Adamic01}, it deals
with a scenario that has not been very exhaustively explored although, in our
opinion, is quite interesting in the communications field: \emph{one-hop
replication networks}.

\paragraph*{One-hop Replication}
One-hop replication networks (also called \emph{lookahead
networks}~\cite{Manku04}) are networks where each node knows the identity of its
neighbors and so it can reply on their behalf. Hence, to find a certain node by
a random walk it suffices to visit any of its neighbors.  This feature is
present for example in social networks, where to find some person it is usually
enough to locate any of her/his friends~\cite{Adamic01}. Also, certain proposals
to improve the resource location process on P2P systems~\cite{Chawathe03,
Cholvi04} (some based on random walks) assume that each node knows the resources
held by its neighbors, so to discover some resource (such as a file or a
service) it suffices to visit any of the neighbors of the node(s) holding it.

\begin{figure}[t]
\begin{center}
\resizebox{0.45\textwidth}{!}{\includegraphics{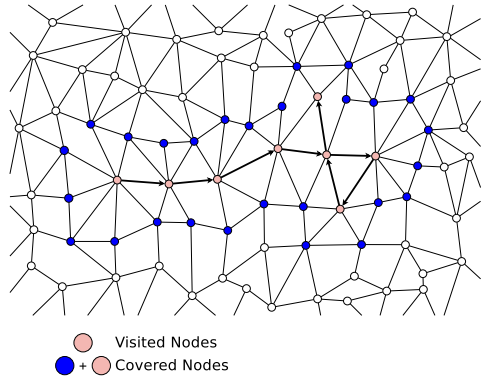}}
\caption{Illustrative example of visited and covered nodes}
\label{fig:illustration}
\end{center} 
\end{figure}

In one-hop replication networks, when the random walk visits some node $i$ we
say it also \emph{discovers} the neighbors of $i$. Hence, we will use two
different terms to refer to the coverage of the random walk.  We denote by
\emph{visited nodes} those that have been traversed by the random walk, and
by \emph{covered nodes} the visited nodes and their neighbors. See
Figure~\ref{fig:illustration} for an illustrative example. 

\paragraph*{Previous Work and the Revisiting Effect} There is some research work
related with the characterization of random walks in one-hop replication
networks.  In~\cite{Mihail07} the authors prove that in the power-law random
graph the amount of hops for a random walk to discover the graph is sublinear
(faster than coupon collection, with which the random walk is compared
in~\cite{Gkantsidis04}).  Also, Manku et al.~\cite{Manku04} study the impact of
lookahead on P2P systems where searches are routed through greedy mechanisms.
In another work, Adamic et al.~\cite{Adamic01} try to find analytical
expressions for $C_G$the cover time of a random walk in power-law
networks with two-hops replication. They detected divergences between the
analytical predictions and the experimental results.  The reason for such
discrepancy, as the authors point out, is the \emph{revisiting effect}, which
occurs when a node is visited more than once. In small-world networks, where a
small number of nodes are connected to other nodes far more often than the rest,
it is quite common for random walks to visit often these highly connected nodes. 

\paragraph*{Our Contributions}
Although there is a plethora of interesting results about random walks, we have
noticed that there are situations where current findings are not straightforward
to apply, especially on communication networks with one-hop replication. For
example, in such networks, we can be interested on studying beforehand the
expected behavior of the random walk to evaluate if it suits the system
requirements.  We characterize the random walk performance by four values:
\begin{itemize}
    \item \emph{The expected coverage}. Given by the expected number of visited
    and covered nodes of each degree $k$ at each hop $l$ of the random walk.
    \item \emph{The expected average search length}. Expected length of searches
    in number of hops, assuming that the source and destination nodes of
    each search are chosen uniformly at random. Obtained from the coverage
    estimations.
    \item \emph{The expected average search duration}. Expected time to solve
    searches. Obtained from the average search length, given the
    \emph{processing capacity} of each node and the \emph{load} on the network
    due to queries.
    \item \emph{The maximum load that can be injected to the network} without
    overloading it.
\end{itemize}

In this work we provide a set of expressions that model the behavior of the
random walk and give estimations for the three previous parameters. Our claim is
that these expressions can be used as a mathematical tool to predict how random
walks will perform on networks of arbitrary degree distribution.  Then, we do
not only address the coverage problem (i.e. to estimate the amount of nodes
covered after each hop of the random walk), but we also apply queuing theory to
model the response time of the system depending on the load. As we show, this
approach allows to compute in advance important magnitudes, such the expected
search duration or the maximum load that can be managed by the network before
getting overloaded. Additionally, we find our model useful to study how certain
features of the network impact on the performance of searches. For example we
find that the best average search time is achieved only if the nodes with higher
degrees have also greater processing capacities.

The expressions related with the estimation of covered nodes at each hop are the
most complex part of the model. They must deal both with the one-hop replication
feature and the revisiting effect. However, we should remark that the model can
be trivially adapted to networks where the \emph{one-hop replication} property
does not hold, and the search finishes only when the node we are searching for
is found (see the last paragraph in Section~\ref{subsec:averageLength}).

Likewise, it is easy to modify the model to a variation of the random walk where
each node avoids sending back the message to the node it received it from at the
previous hop. We denote this routing mechanism \emph{avoiding random walks}, and
we deem it interesting for two reasons. First, intuitively, it should improve
the random walk coverage (we have confirmed this experimentally). Second, it can
be implemented in real systems using only local information, just as the pure
random walk (the sending node only needs to know from which neighbor the message
came from).
 
A feature of our proposal is that it does not require the complete adjacency
matrix $\mathbi{A}$, that in some situations could be unknown.  Instead, thanks
to the randomness assumption we apply it only needs the degree distribution of
the network to compute the metrics we are interested in.  On the other hand,
this work is focused on networks with good connectivity and where the nodes
degrees are independent (see Section~\ref{subsec:modelAndAssumptions}).

Another property of this model is that it takes into account the revisiting
effect by modeling the coverage of the random walk at each hop $l$ depending on
the coverage at the previous hop $l-1$.  That is, the evolution of the coverage
is not assumed to be a memoryless process, a simplification that can lead to
errors as seen in~\cite{Adamic01}.

The rest of the paper is organized as follows.
Section~\ref{sec:knownNodesAverLeng} introduces our analysis of the coverage and
average search length of random walks, along with some experimental evaluation.
Section~\ref{sec:searchesByRW} is centered on obtaining the average search time
of random walks. Finally, in Section~\ref{sec:conclusions}, we state our
conclusions and propose some potential future work.


\section{Analysis of Random Walks}
\label{sec:knownNodesAverLeng}

In this section, we analyze the behavior of random walks in arbitrary networks.

\subsection{Model and Assumptions}
\label{subsec:modelAndAssumptions}
We will represent networks by means of undirected graphs $G = (V , E)$, where
vertices $V$ represent the nodes and edges $E\subseteq V \times V$ are the links
between nodes. There are no links connecting a vertex to itself, or multiple
edges  between the same two vertices. This does not simplify our model, but
makes it closer to real scenarios like typical P2P networks. We denote by
$|V|=n$ the number of nodes in the graph and by $n_k$ the number of nodes that
have degree $k$ (i.e., the number of nodes that have $k$~neighbors,
$\sum_kkn_k=2|E|$).  For all vertices its degree $k$ is lower than the size of
the network $n$, as in typical real world networks (such as social and pure P2P
networks) each node is connected to only a subset of the other vertices in the
system\footnote{Some P2P networks like Napster have a central node that network
members use to locate files. But those networks are not considered as pure P2P
systems because they use a typical server-client architecture with a centralized
topology to perform searches. They are regarded to have a ``P2P'' behavior only
in the way files are shared. This work is rather focused on the decentralized
topologies of pure P2P networks}. We also denote by $p_k$ the probability that
some node in the network, chosen uniformly at random, has degree $k$ (i.e.,
$p_k=n_k/n$). The average degree of a network is given by $\overline{k}=\sum_k k
\; p_k$. For a given network, the distribution formed by the probabilities $p_k$
(for all $k$) is known as the \emph{degree distribution} of such a network.


A random walk over $G$ can be defined as a \emph{Markov Chain}~\cite{Motwani95}
process $M_G$ where the transition matrix $\mathbi{P}=[P_{ij}]$ is defined as:
\begin{equation}
    P_{ij}\:=\:\begin{cases}
                 \frac{1}{d(i)}\:\:\text{if}\:(i,j)\in E.\\
                 0\:\:\:\:\:\:\text{otherwise.}
               \end{cases}
\end{equation}

where $P_{ij}$ is the probability of moving from node $i$ to node $j$, and
$d(i)$ is the degree of node $i$. $\mathbi{P}$ allows to study the probability
of visiting each node at each hop $l$. This probability is expressed in the
\emph{state probability vector}, $\mathbi{q}^l=(q_1^l, q_2^l, ...,
q_n^l)$, where $q_i^l$ represents the probability that the random walk visits
node $i$ at hop $l$. This probability evolves as
$\mathbi{q}^l=\mathbi{q}^{l-1}\mathbi{P}$.

Assuming that $G$ is connected and finite, then $M_G$ is irreducible: any node
can be reached from any other node, and the average path length between two any
nodes is finite. Assuming also that $G$ is non-bipartite, then we can state that
$M_G$ is aperiodic and so we are able to apply the \emph{Fundamental Theorem of
Markov Chains}~\cite{Motwani95}. This theorem states that in such graph $M_G$ is
\emph{ergodic} an exists an unique state probability distribution
$\boldsymbol{\pi}$, denoted the \emph{stationary distribution}, such that
$\boldsymbol{\pi}\mathbi{P}=\boldsymbol{\pi}$, $\pi=(\pi_1, \pi_2, ... ,\pi_n)$,
where $\pi_i$ is:

\begin{equation}
    \pi_i\:=\:\frac{d(i)}{2|E|}.
\end{equation}

Intuitively, $\boldsymbol{\pi}$ represents the steady state of $M_G$. That is,
$\pi_i$ represents the probability that the node $i$ is visited at any hop of
the random walk once the stationary distribution has been reached. This
probability is proportional to the degree of $i$, $d(i)$.  

\paragraph*{Mixing Rate and Conductance}
We are interested on how fast the random walk converges to $\boldsymbol{\pi}$, a
magnitude that is called the \emph{mixing rate}~\cite{Lovasz93}. We require a
fast convergence in order to be able to apply Equation~\ref{equ:Pa}.

The convergence rate is related with the eigenvalues of the transition matrix
$\mathbi{P}$. A vector $\vec{x}$ is an \emph{eigenvector} of $\mathbi{P}$ with
\emph{eigenvalue} $\lambda$ iff $\vec{x}\mathbi{P}=\lambda\vec{x}$ (so for
example $\boldsymbol{\pi}$ is an eigenvector of $\mathbi{P}$ with eigenvalue
$1$). It is well known~\cite{Lovasz93} that $\mathbi{P}$ has $n$ real
eigenvalues $\lambda_0=1>\lambda_1\geq...\geq\lambda_{n-1}\geq-1$ (and in fact,
if $G$ is non-bipartite then $\lambda_{n-1}>-1$). It is also
known~\cite{Sinclair92} that the convergence rate to $\boldsymbol{\pi}$ is
governed by the second largest eigenvalue modulus of $\mathbi{P}$, $\max
\{\lambda_1,|\lambda_{n-1}|\}$. In most real world networks we can safely assume
that $\lambda_1>|\lambda_{n-1}|$~\cite{Lovasz93,Sinclair92,Gkantsidis04}. The
following holds for a random walk starting at node $i$~\cite{Lovasz93}:
\begin{equation}
    |\mathbi{P}_i^{(l)}(j) - \boldsymbol{\pi}_j| \leq
    \sqrt{\frac{d(j)}{d(i)}}\lambda_1^l,
\end{equation}

where $\mathbi{P}_{i}^{(l)}$ is the distribution of the state of the random walk
at hop $l$, when $i$ is the initial state. Thus, we can expect a fast mixing
for high values of the \emph{spectral gap} $1 - \lambda_1$.

Now, the $\lambda_1$ value is strongly related with the \emph{conductance} of
the network, $\Phi_G$. Informally, the conductance measures how well `connected'
the graph is. It is defined as follows. For $S\subseteq V$, the cutset of $S$,
$C(S)$, is the set of edges with one endpoint in $S$ and the other endpoint in
$\Bar{S}$.  The volume of $S$, $\text{vol}(S)$, is defined as the sum of degrees
of the nodes in $S$, i.e., $\text{vol}(S)=\sum_{i \in S}d(i)$. Then the
conductance of $G$ is computed as:

\begin{equation}
    \Phi_G = \min_{\substack{
    S\subset V \\
    \linebreak \text{vol}(S)\leq \text{vol}(V)/2
    }}\frac{|C(S)|}{\text{vol}(S)}.
\end{equation}

The relationship between the conductance and the convergence is given by the
following expression (\emph{Cheeger's inequality})~\cite{Lovasz93}:
\begin{equation}
    \frac{\Phi_G^2}{2}\leq 1 - \lambda_1 \leq 2\Phi_G.
\end{equation}

So \emph{a good conductance leads to high mixing rates}, that is, the random
walk state will converge quickly to the stationary distribution
$\boldsymbol{\pi}$. The intuition behind this fact is that in graphs with good
conductance the random walk will be able to move to any region of the graph
easily, whichever the origin node, and so it will evolve quickly to the
equilibrium. We reason that high connectivity is to be expected in many real
world networks (specially communication networks) and network
models~\cite{Tahbaz07,Gkantsidis03,Broder00}.

Therefore, we can assume that the probability that the node visited by the
random walk has degree $k$ at each hop of the random walk, $P(k)$, is also
proportional to $k$ and can be computed as:

\begin{equation}
\label{equ:Pa}
    P(k)\:=\:
    \sum_{\substack{i\in V \\ \: d(i)=k}}\frac{d(i)}{2|E|}\:=\:
    n_k \frac{k}{\sum_j jn_j}\:=\:\frac{k \; p_k}{\overline{k}}.
\end{equation}

We will apply Equation~\ref{equ:Pa} intensively for our analysis of the
coverage. Of course, its correctness depends on the distance of the random walk
to the stationary distribution, or how fast it converges to it. Another
issue to be taken into account is the possible dependencies between successive
steps of the random walk. Our analysis estimates the average number of nodes
visited and covered by the random walk at a certain hop from the values
estimated at the previous hop. The new estimation is done assuming that the
random walk has statistical properties similar to the random sampling of nodes
where the probability of choosing a certain node is proportional to $k_i$,
\emph{despite the apparent dependencies between consecutive hops}. 

Also, the work by Gkantsidis et al.~\cite{Gkantsidis04} shows the similarities
between independent sampling and random walks, that we assume for our mean based
analysis. As the authors state, in networks with good connectivity and expansion
properties (which are strongly related to $\lambda_1$) the random walk has a
behavior close to independent sampling, being the probability of choosing some
node proportional to its degree.

\begin{figure*}[t]
\begin{center}
\subfigure[Erdos-Renyi networks.]{
\resizebox{0.45\textwidth}{!}
    {\includegraphics{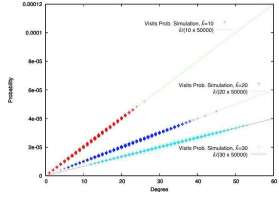}}
\label{fig:ER}
}
\subfigure[Small-world networks.]{
\resizebox{0.45\textwidth}{!}
    {\includegraphics{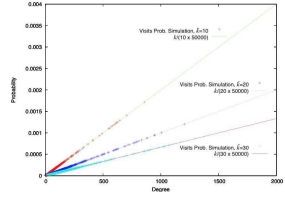}}
\label{fig:Newman}
}
\caption{In these figures, we show the probability of a search message arriving
at a particular node as a function of its degree. We have used both Erdos-Renyi
and small-world (power-law) networks formed by $50,000$ nodes, with different
average node degrees ($10$, $20$ and $30$). The same experiments have been
performed with networks formed by $25,000$ and $100,000$ nodes, and we found
similar results.  As it can be readily seen, the probability of a search message
arriving at a particular node is proportional to the degree of the node.}
\label{fig:property}

\end{center} 
\end{figure*}

Besides, we have performed some experiments to verify the correctness of this
hypothesis.  The results, shown in Figure~\ref{fig:property} confirm it is a
valid assumption. Also, we would like to remark that the property expressed by
Equation~\ref{equ:Pa} is in fact assumed in previous works about random walks
(e.g.,~\cite{Newman01,Adamic01}) and backed by~\cite{Gkantsidis04}.

\begin{figure*}[t]
\begin{center}
\subfigure[Erdos-Renyi network, $\overline{k}=30$.]{
\resizebox{0.47\textwidth}{!}
    {\includegraphics{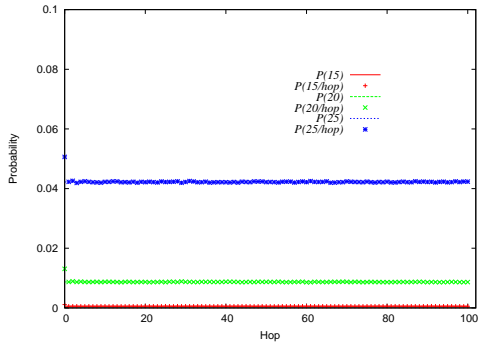}}
\label{fig:ERIndHop}
}
\subfigure[Small-world network, $\overline{k}=10$.]{
\resizebox{0.47\textwidth}{!}
    {\includegraphics{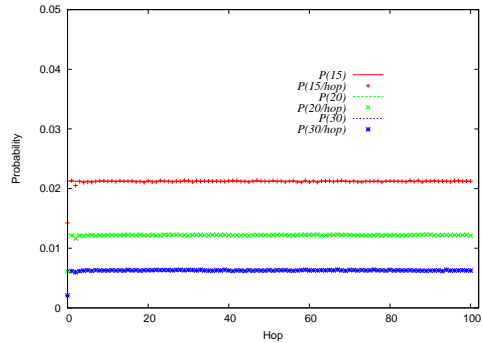}}
\label{fig:NewmanIndHop}
}
\caption{These figures compare the probability $P(k)$ of reaching a node of
degree $k$ as defined by the model, with the measured probability of reaching a
node of degree $k$ at each hop of the random walk. Both for ER and small-world
networks the experimental results are averaged over three different networks
with the same average degree and size ($n=50\cdot 10^4$).}
\label{fig:property3}

\end{center} 
\end{figure*}

Another important issue we have tested is how `fast' the random walk evolves to
a state where the assumption of Eq.~\ref{equ:Pa} holds.
Figure~\ref{fig:property3} shows how the random behaves. It can be seen that,
almost immediately after hop 0 (start node), the probability of reaching a node
of degree $k$ is $P(k)$.

We should note that the good conductance property, that implies that the random
walk can move from any node to any other node in few steps, discards some
topologies such as cycles.

\begin{figure*}[t]
\begin{center}
\subfigure[Erdos-Renyi network, $\overline{k}=30$.]{
\resizebox{0.47\textwidth}{!}
    {\includegraphics{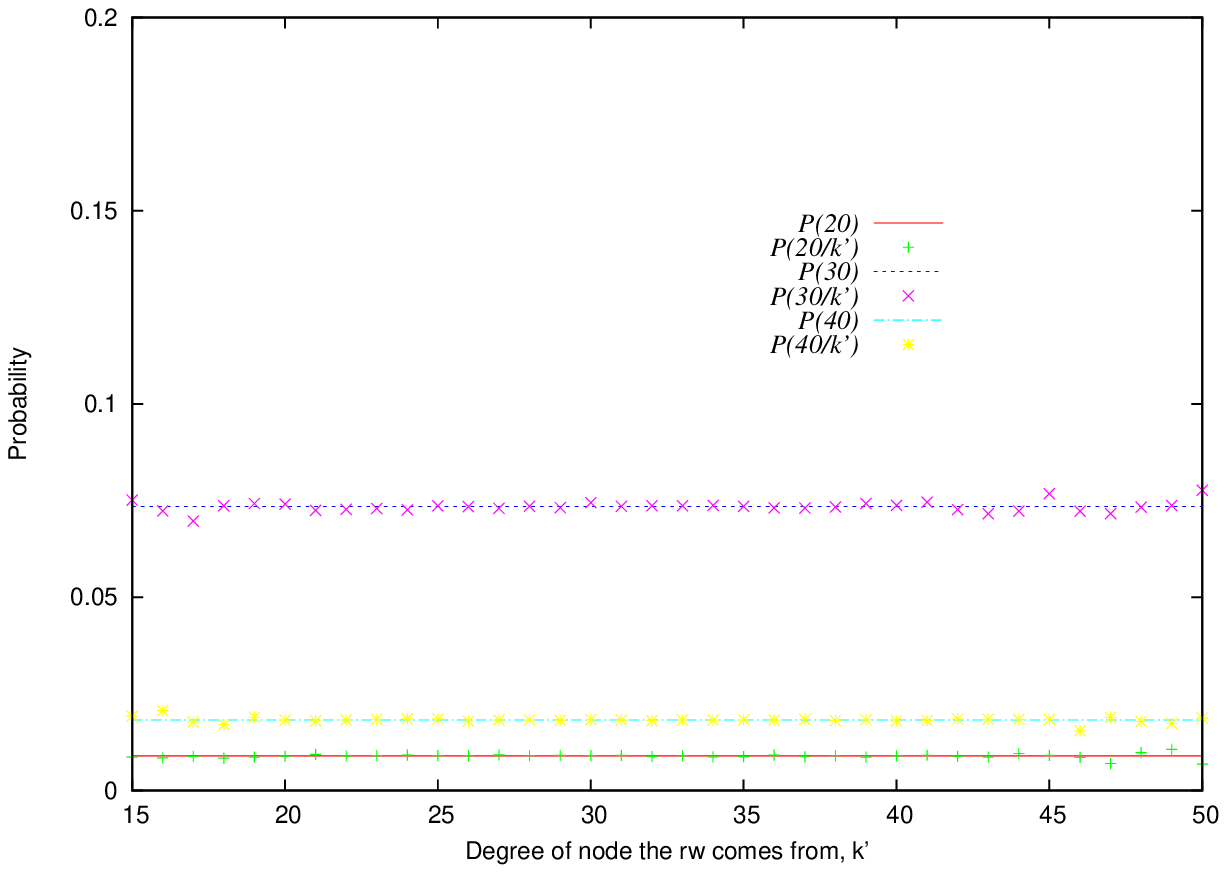}}
\label{fig:ERIndDeg}
}
\subfigure[Small-world network, $\overline{k}=10$.]{
\resizebox{0.47\textwidth}{!}
    {\includegraphics{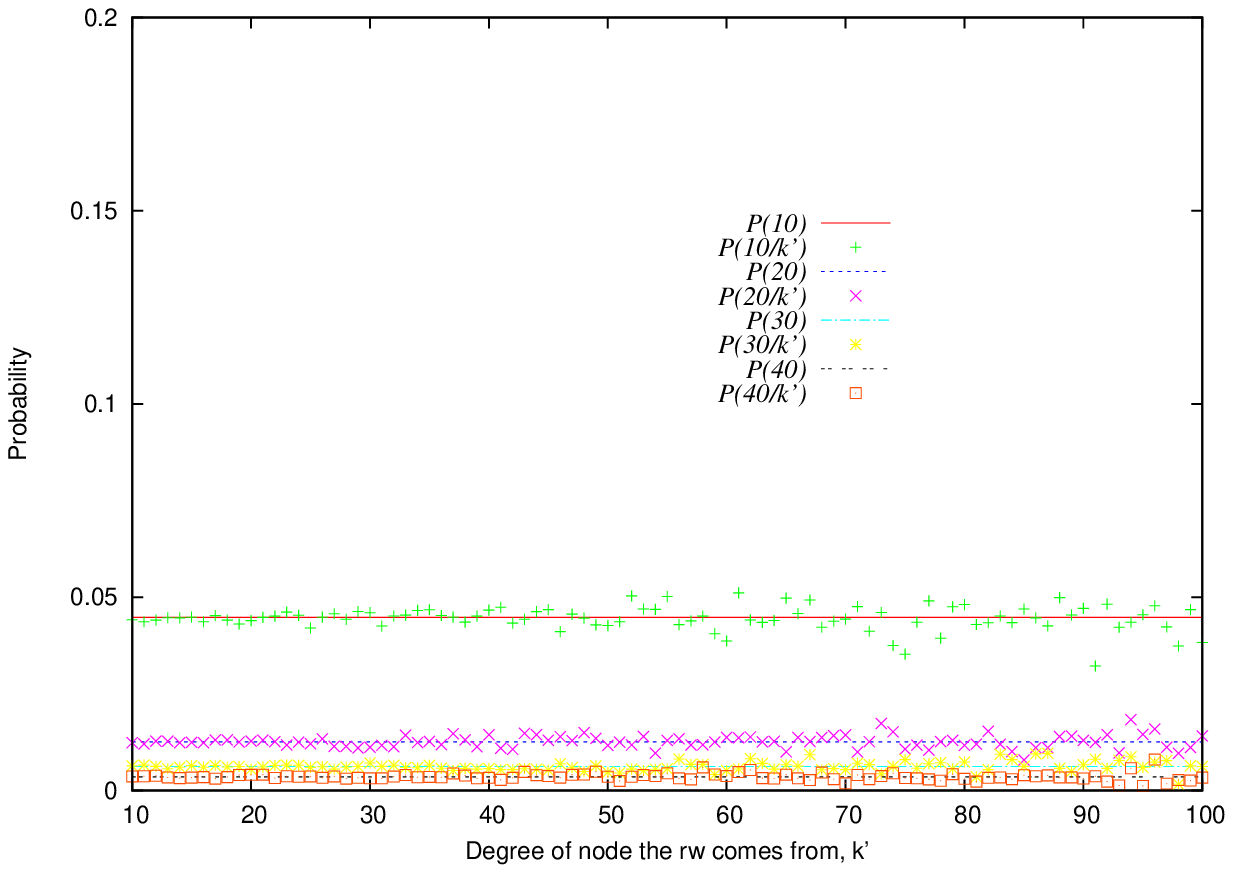}}
\label{fig:NewmanIndGed}
}
\caption{These figures compare the probability $P(k)$ of reaching a node of
degree $k$ as defined by the model, with the measured probability of reaching a
node of degree $k$ given that the rw comes from a node of degree $k'$,
$P(k/k')$. Both for ER and small-world networks the experimental results are
averaged over three different networks with the same average degree and size
($n=10^5$).}
\label{fig:property2}

\end{center} 
\end{figure*}

\paragraph*{Independence of Nodes Degrees} 
Finally, we assume that the degrees of neighbors are independent. That is, given
any two connected nodes $i$ and $j$ ($(i,j)\mathbin{\in}E$) and any two degree
values $k_1$ and $k_2$, then
$P[d(i)\mathopen{=}k_1\mathbin{|}d(j)\mathopen{=}k_2]=P[d(i)\mathopen{=}k_1]=p_{k_1}$.
This property holds in networks built by random mechanisms, like the ones used
to built the ER and small-world networks we target in our experiments. To
confirm that the degree independence assumption is valid we have run some
experiments, whose results are shown in Figure~\ref{fig:property2}. These
experiments aim to measure if the probability of reaching a node of degree $k$
when following a random walk is affected by the degree $k'$ of the node the
random walk was in the previous hop ($P(k/k')$). Our results lead to the
conclusion that $\forall k,k' P(k/k')=P(k)$, that is, $k'$ does not have an
impact on $k$.

We should note also that this property is not fulfilled in certain graphs like
those built by preferential mechanisms where it is well-known that there is a
correlation among neighbors degrees~\cite{Krapivsky01}.  This could lead to
certain deviations in mean-based analysis of the random walk (as our own).

In the following, we study how many different nodes are visited by a random walk
as a function of its length (i.e., of the number of steps taken) and of the
degree distribution of the chosen network. Subsequently, we extend this result
to also consider the neighbors of the visited node. These metrics allow us to
quantify how much of a network is being ``known'' throughout a random walk
progress.  Then, we turn our attention to provide an estimation of the average
search length of a random walk. In the last subsection, we validate our
analytical results by means of simulations. We assume that only the degree
distribution $p_k$ and the size $n=|V|$ of the network are known.

\subsection{Number of Visited Nodes}
\label{subsec:visitedNodes}

This metric represents the average number of different nodes that are visited by
a random walk until hop $l$ (inclusive), denoted by $V^l$. Note that nodes may
each be visited more than once, but revisits are not counted.

To obtain $V^l$, we first calculate the average number of different nodes of
degree~$k$ that are visited by a random walk until hop $l$ (inclusive), denoted
by $V_k^l$. We make a case analysis:

\begin{itemize}

\item
When $l=0$ (i.e., in the source node):  Since the source node of the random walk
is chosen uniformly at random, then the probability of starting a random walk at
a node of degree $k$ is $p_k$. Therefore, 

\begin{equation}
\label{eq:visited0}
    V_k^0\:=\:1\cdot p_k\:+\:0\cdot (1-p_k)\:=\:p_k.
\end{equation}

\item
When $l=1$ (i.e., at the first hop): Here we apply that the probability of
visiting some node of degree $k$ at any hop is given by $P(k)$
(Equation~\ref{equ:Pa}). This is based on the assumption that the random walk
behaves similarly to independent sampling despite dependencies between
consecutive hops (based on~\cite{Gkantsidis04}, see
Section~\ref{subsec:modelAndAssumptions}). We deem this premise to be reasonable
even at the first stages of the random walk, due to the high mixing rates found
in the type of networks on which we focus our work (again, see
Section~\ref{subsec:modelAndAssumptions}). Recall that the experimental
evaluation both of this assumption (Fig.~\ref{fig:property}) and of our model
(shown in Section~\ref{sec:experimentalResults}), seem to verify this. Thus, we
have that 

\begin{equation}
\label{eq:visited1}
\begin{split}
    V_k^1\: & =\:V_k^0+P(k) \\
    & = p_k + \frac{k \; p_k}{\overline{k}}.
\end{split}
\end{equation}

\item
When $l>1$: we must take into account the probability of the random walk
arriving at an already visited node. To compute such a probability, we define
the following two values:

\begin{itemize}
    \item $P_v(k,l)$: This represents the probability that, if the random walk
    arrives at a node of degree $k$ at hop $l$, that node has been visited
    before. It can be obtained as follows:
     \begin{equation}
        P_v(k,l)\:=\:\frac{V_k^{l-2}}{n_k}.
    \end{equation}
    
    Note that we put $V_k^{l-2}$ instead of $V_k^{l-1}$ because the node visited
    at hop $l-1$ can not be visited at hop $l$ (no vertex is connected to
    itself).
        
     \item $P_b$: This is the probability that at any given hop the random walk
     is moving back to the node where it came from\footnote{Here we can easily
     adapt the model to the \emph{avoiding random walk}.  If we don't want to
     consider the case of a random walk moving back to the node where it came
     from, it is enough to assign $P_b=0$.}.  Since any visited node has degree
     $k$ with probability $P(k)$, then the random walk will go back through the
     same link from which it came with probability $1/k$. Therefore, we have:

    \begin{equation}
    \label{equ:probBack}
        P_b\:=\:\sum_k P(k)\frac{1}{k}\:=\:\frac{1}{\overline{k}}.
    \end{equation}
    
\end{itemize}

Using these probabilities, $V_k^l$ can be written as

\begin{equation}
\label{eq:visitedn}
\begin{split}
    V_k^l \; & = \; V_k^{l-1}+P(k)(1-P_b)(1-P_v(k,l)) \\
    & = \; V_k^{l-1}+\frac{k \; p_k}{\overline{k}}
    \left(1-\frac{1}{\overline{k}}\right)  \left(1-\frac{V_k^{l-2}}{n_k}\right).
\end{split}
\end{equation}

\end{itemize}

Finally, taking the results obtained in
Equations~\ref{eq:visited0},~\ref{eq:visited1}  and~\ref{eq:visitedn}, we have
that the total number of different nodes visited until hop $l$ is
\begin{equation}
    V^l\:=\:\sum_k V_k^l.
\end{equation}


\subsection{Number of Covered Nodes}
\label{subsec:knownNodes}

This metric provides an estimation of the average number of different nodes
\emph{covered} by a random walk until hop~$l$ (inclusive), denoted by $C^l$. A
node is covered by a random walk if such a node, or any of its neighbors, has
been visited by the random walk.

To obtain $C^l$, we first calculate the number of different nodes of degree~$k$
covered at hop~$l$, denoted by $C_k^l$.

\begin{itemize}
\item
When $l=0$:
\begin{equation}
\label{eq:covered0}
\begin{split}
    C_k^0\: & =\:p_k (1 + k P(k))\:+\:\sum_{j\neq k} p_j \; j \; P(k) \\
    & =\:V_k^0+P(k)\;\overline{k}.
\end{split}
\end{equation}

The first term takes into account the possibility that the source node has
degree~$k$. The second term refers to the number of neighboring nodes (of the
source node) of degree~$k$. If the source node has degree~$j$ (which happens
with probability $p_j$) then, on average, $j \; P(k)$ nodes of degree $k$ will
be covered, since each one of the $j$ neighboring nodes of the source node will
have degree~$k$ with probability $P(k)$.

\begin{figure}[t]
\begin{center}
\resizebox{0.45\textwidth}{!}{\includegraphics{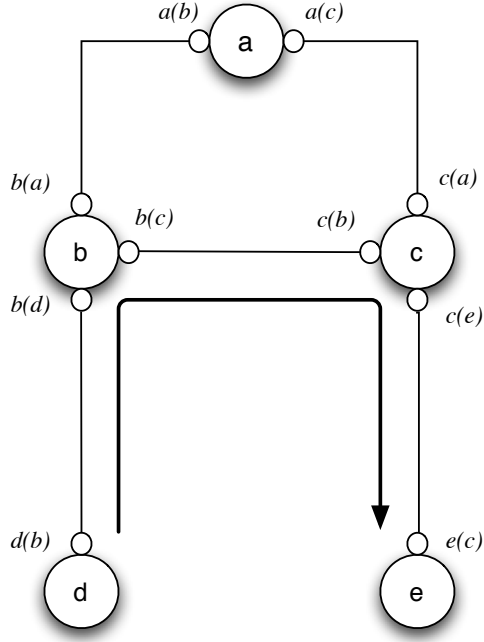}}
\caption{The figure shows a simple graph formed by 5 nodes (named $a$, $b$, $c$,
$d$ and $e$) where there is a random walk that follows the path $d-b-c-e$.  At
each node, we represent the different "endpoints" that are \emph{hooked} on that
node by means of small circles. For instance, the endpoints $a(b)$ and $a(c)$
are said to be hooked onto node $a$.  In the graph, when the random walk starts
(at node $d$), then endpoint $b(d)$ is said to be \emph{checked}. Similarly,
when it visits node $b$, then endpoints $d(b)$, $a(b)$ and $c(b)$ are said to be
checked. The same mechanism applies when the random walk visits nodes $c$ and
$e$.}
\label{fig:endpoints}
\end{center}
\end{figure} 

\item
When $l>0$:  Given a link $(\varv,\varw) \in E$, we say that it has two
endpoints, which are the two ends of the link.  We denote the endpoint of the
link at node~$\varv$ by $\varv\,(\varw)$, and similarly the endpoint of the link
at node~$\varw$ by $\varw\,(\varv)$. We say that $\varv\,(\varw)$ \emph{hooks
onto} node~$\varv$. We also say that $\varv\,(\varw)$ has been \emph{checked} by
a random walk if such a random walk has visited node $\varw$. These concepts are
graphically explained in Fig.~\ref{fig:endpoints}.

Now, let us denote by $E^l$ the number of endpoints checked for the first time
at hop~$l$, and by $P_{u}(k,l)$ the probability that these endpoints hook onto
still uncovered nodes of degree~$k$. Then, $C_k^l$ (where $l>0$) can be written
as follows:

\begin{equation}
\label{eq:uno}
    C_k^l\:=\:C_k^{l-1}\:+\:  P_{u}(k,l) \;E^l.
\end{equation}

\begin{itemize}
\item
To obtain $E^l$, we consider the number of different endpoints checked after
hop~$l$ to be $\sum_j j V_j^l$. So, the number of endpoints checked for the
first time at hop~$l$ is $\sum_j (V_j^l - V_j^{l-1}) j$. However, one of the
endpoints hooks onto the node the random walk comes from (i.e., it cannot
increase the amount of nodes that are covered). Thus:

\begin{equation}
\label{eq:dos}
    E^l = \sum_j (V_j^l - V_j^{l-1}) (j-1).
\end{equation}

\item
To obtain $P_{u}(k,l)$, on one hand we consider the overall number of endpoints
hooking onto uncovered nodes of degree~$k$ just before hop~$l$ is $k (n_k -
C_k^{l-1})$. On the other hand, the overall number of endpoints is $\sum_j j\;
n_j$, and the overall number of checked endpoints until hop~$l-1$ (inclusive) is
$\sum_j j \; V_j^{l-1}$. That is, the number of endpoints not checked just
before hop~$l$ is $\sum_j j \; n_j - \sum_j j \; V_j^{l-1}$.  Therefore, we can
write:

\begin{equation}
\label{eq:tres}
    P_{u}(k,l)= \frac{k\; (n_k-C_k^{l-1})}{\sum_j j \; n_j - \sum_j j \;
    V_j^{l-1}}.
\end{equation}

\end{itemize}

Substituting Equation~\ref{eq:dos} and~\ref{eq:tres} into Equation~\ref{eq:uno},
we have that 

\begin{equation}
\label{eq:coveredn}
\begin{split}
    C_k^l\:=\: & C_k^{l-1}\:+\:
    \left(\frac{k\; (n_k-C_k^{l-1})}
    {\sum_j j \; n_j - \sum_j j \; V_j^{l-1}} \right)
    \times \sum_j (V_j^l-V_j^{l-1})\:(j-1).
\end{split}
\end{equation}

\end{itemize}

Finally, taking into account Equations~\ref{eq:covered0} and \ref{eq:coveredn},
we have that the total number of nodes covered after hop $l$ is
\begin{equation}
    C^l\:=\:\sum_k C_k^l.
\end{equation}


\begin{figure}[t]
\begin{center}
\subfigure[Erdos-Renyi network.]{
\resizebox{0.3\textwidth}{!}{\includegraphics{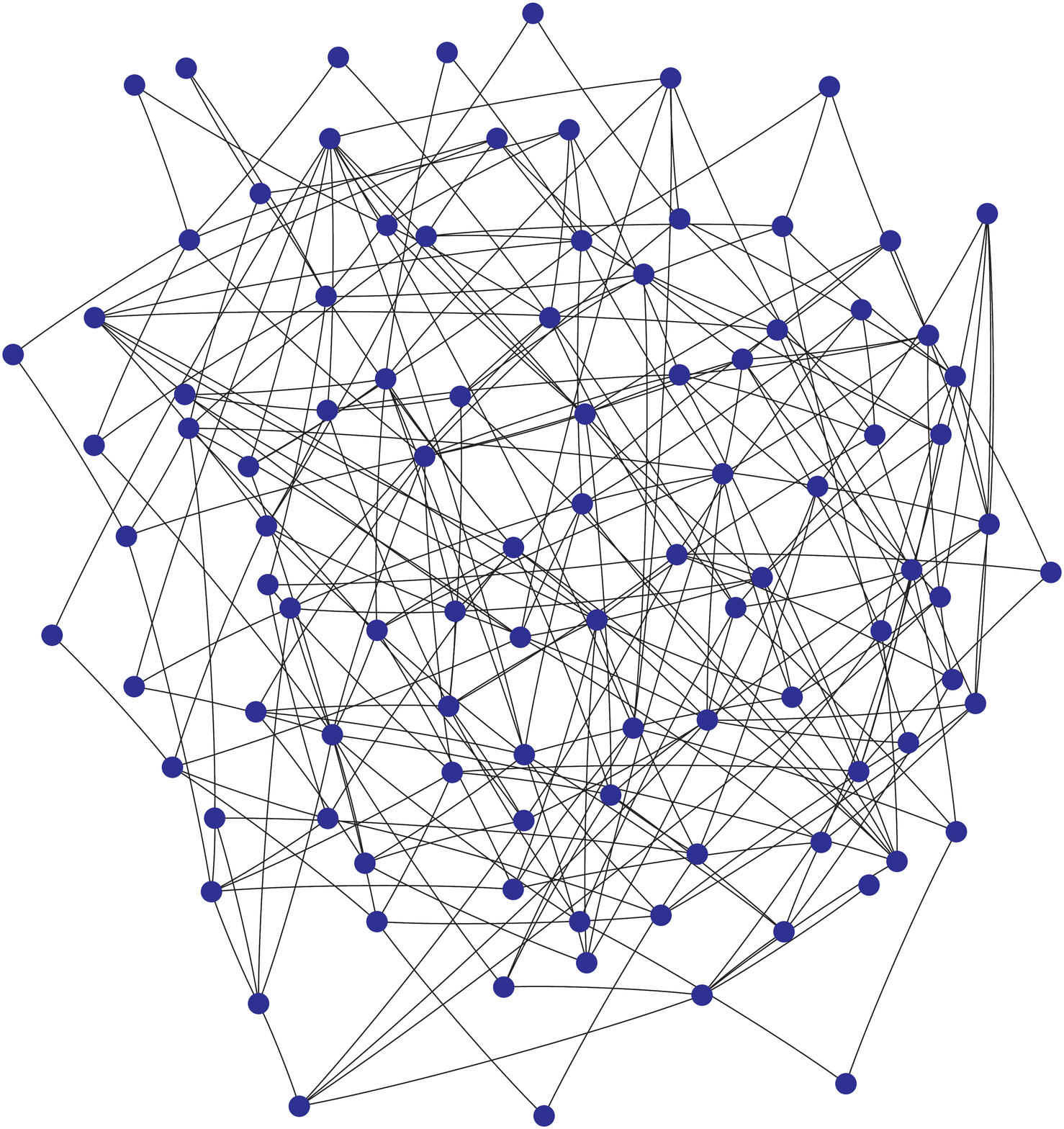}}
}
\hspace{0.5cm}
\subfigure[Small-world network.]{
\resizebox{0.3\textwidth}{!}{\includegraphics{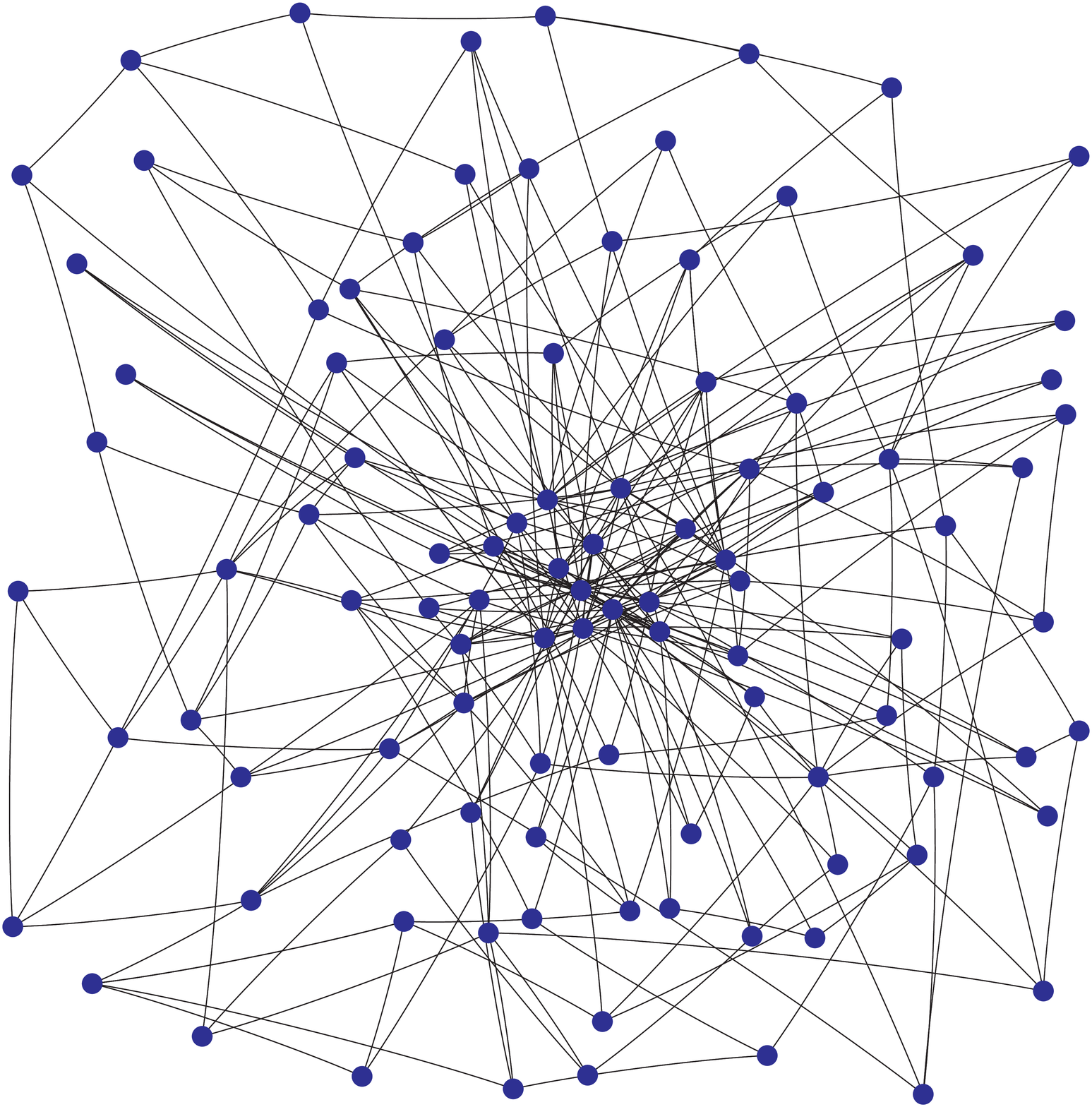}}
}
\caption{In the Erdos-Renyi network most nodes have approximately the same
number of links. In contrast, the small-world network is heterogeneous: the
majority of the nodes have approximately the same number of links but a few
nodes have a large number of them.} 
\label{fig:networks}
\end{center} 
\end{figure} 

\subsection{Average Search Length}
\label{subsec:averageLength}

Using the previous metric, we are now able to provide an estimation of the
average search length of  random walks, denoted by $\overline{l}$.  Formally,
$\overline{l}$ is given by the following expression:
\begin{equation}
\label{equ:lenght}
    \overline{l}\:=\:\sum_{l=0}^{\infty} l \;P_f(l),
\end{equation}

where $P_f(l)$ is the probability that the search finishes at hop $l$ (i.e., the
probability that the search is successful at hop~$l$, having failed during the
previous $l-1$ hops). Let us define the \emph{probability of success} at hop
$l$, denoted by $P_s(l)$, as the probability of finding, at that hop, the node
we are searching for. $P_s(l)$ can be obtained as the relation between the
number of new nodes that will be covered at hop $l$, and the number of nodes
that are still uncovered at hop $l$. That is,

\begin{equation}
\label{equ:successProb}
    P_s(l)\:=\:\frac{C^l-C^{l-1}}{n-C^{l-1}}.
\end{equation}

Now, $P_f(l)$ can be obtained as follows:

\begin{equation}
\label{equ:finishedProb}
    P_f(l)\:=\:P_s(l) \prod_{i=0}^{l-1}{(1-P_s(i))} \:=\:\frac{C^l-C^{l-1}}{n}.
\end{equation}

Therefore, $\overline{l}$ can be written as

\begin{equation}
\label{equ:averageLength}
    \overline{l}\:=\:\frac{1}{n}\sum_{l=0}^{\infty} l \; (C^l-C^{l-1}).
\end{equation}


\subsection{Experimental Evaluation}
\label{sec:experimentalResults}
We have run a set of experiments to evaluate the accuracy of the expressions
presented in the previous subsections. The results obtained are presented in
this section.


For our work, we consider two kinds of network: small-world networks
(constructed as in~\cite{Newman01}) and Erdos-Renyi networks (constructed as
in~\cite{Bollobas01}).
\begin{itemize}
    \item \emph{Small-world networks}~\cite{Newman01,Barabasi99}.
    In~\cite{Albert02} it is shown that many real world networks present an
    interesting feature: each node can be reached from any other node in few
    hops. These networks are typically denoted small-world networks.  The
    Internet, the Web, the Science collaboration graph, etc. are examples of
    real world networks that are consistent with this property. This kind of
    networks are also specially interesting for our work because here the
    revisiting effect commented in Section~\ref{sec:intro} is strongly present
    due to the uneven degree distribution. We build small-world networks using
    the mechanism described in~\cite{Newman01}, which leads to networks whose
    degree distribution follows a power-law distribution $p_k\sim k^{-\alpha}$
    (power-law networks).
    \item \emph{Erdos-Renyi (ER) random networks}~\cite{Bollobas01}. For two any
    nodes $i,j \in V$ there is a constant probability $c$ that they are
    connected $((i,j) \in E$). The resulting degree distribution is a binomial
    distribution $p_k\sim \binom{n}{k}c^k(1-c)^{n-k}$.
\end{itemize}
See Figure~\ref{fig:networks} for an illustrative example of both kinds of
networks.



\begin{figure*}[t]
\begin{center}
\subfigure[Erdos-Renyi and Small-world; $n=5\cdot 10^4$; $\overline{k}=10$,
$30$.]{
\resizebox{0.47\textwidth}{!}{\includegraphics{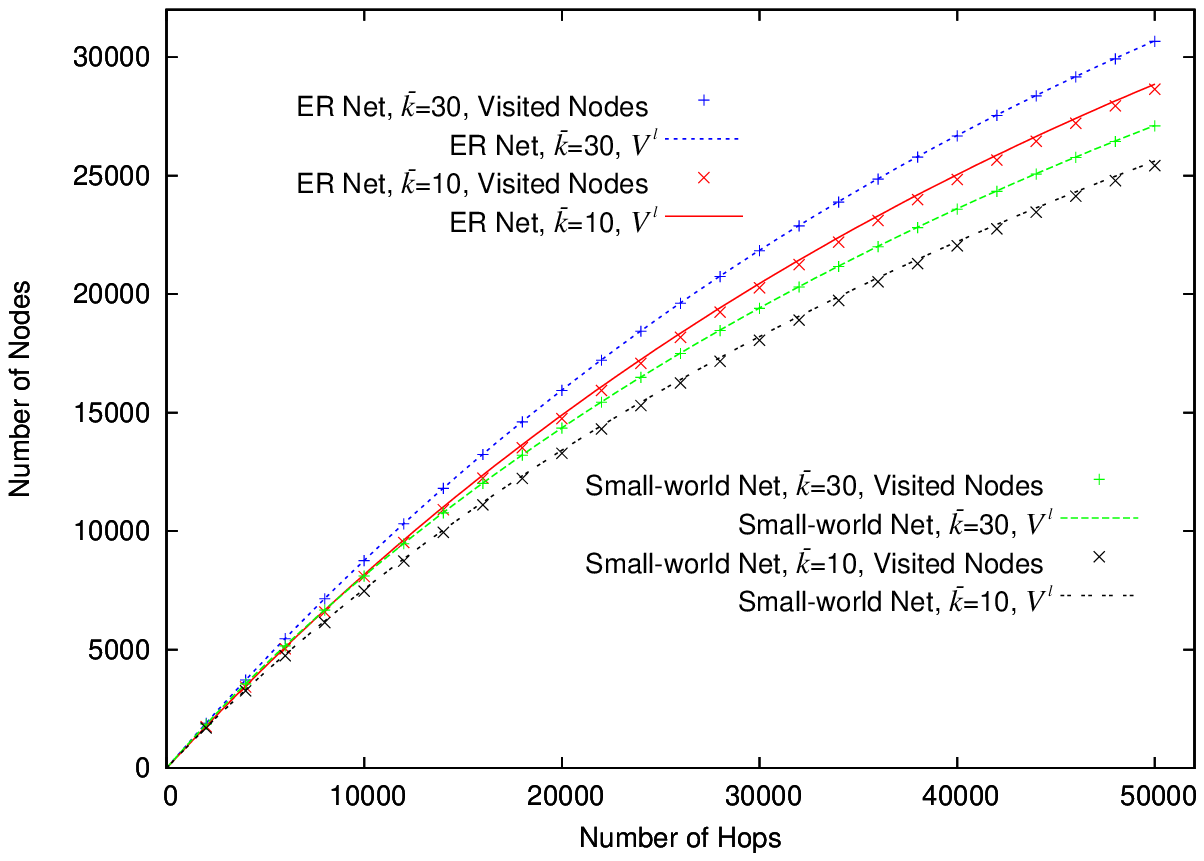}}
\label{fig:visitedRandomSmallWorld50}
}
\subfigure[Erdos-Renyi and Small-world; $n=5\cdot 10^4$, $10^5$;
$\overline{k}=20$.]{
\resizebox{0.47\textwidth}{!}{\includegraphics{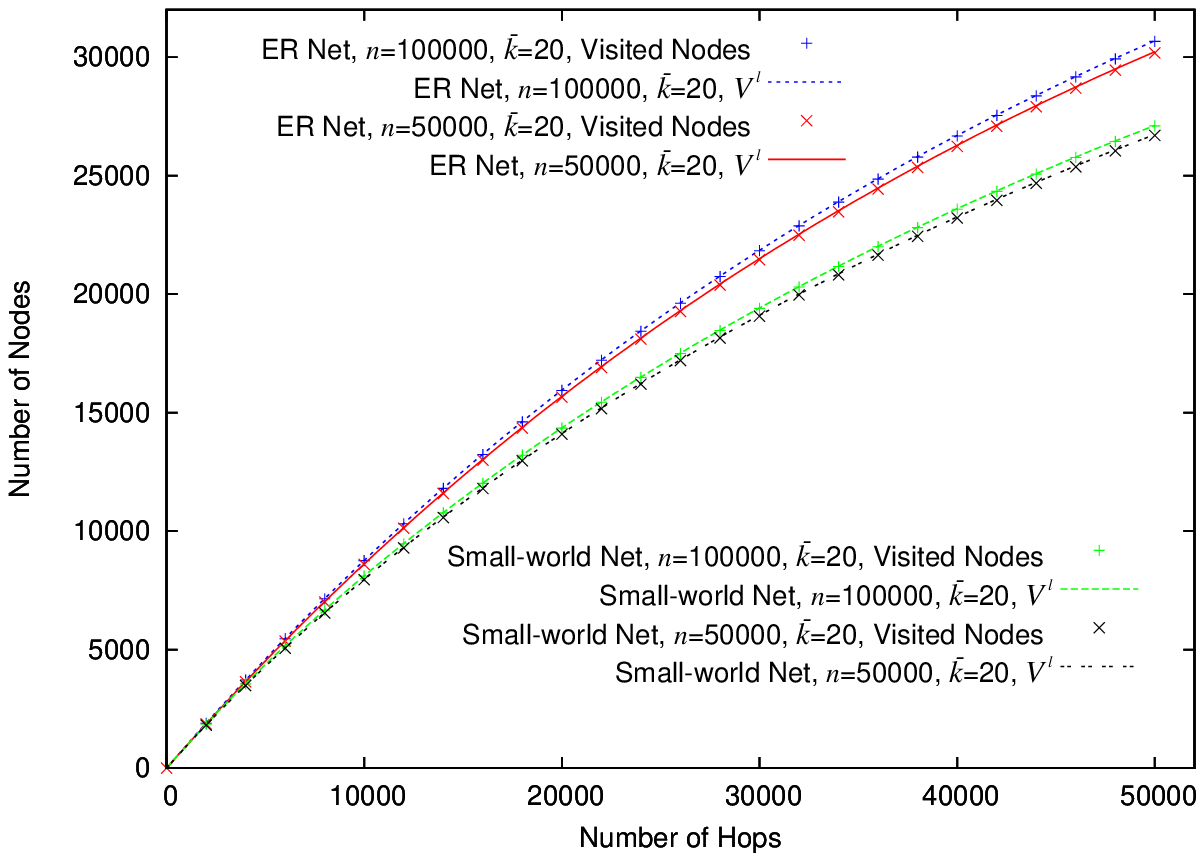}}
\label{fig:visitedRandomSmallWorld50100}
}
\caption{Visited nodes $V^l$.}
\label{fig:visited}
\end{center} 
\end{figure*}

\begin{figure*}[t]
\begin{center}
\subfigure[Erdos-Renyi; $n=10^5$; $\overline{k}=10$, $20$, $30$.]{
\resizebox{0.47\textwidth}{!}{\includegraphics{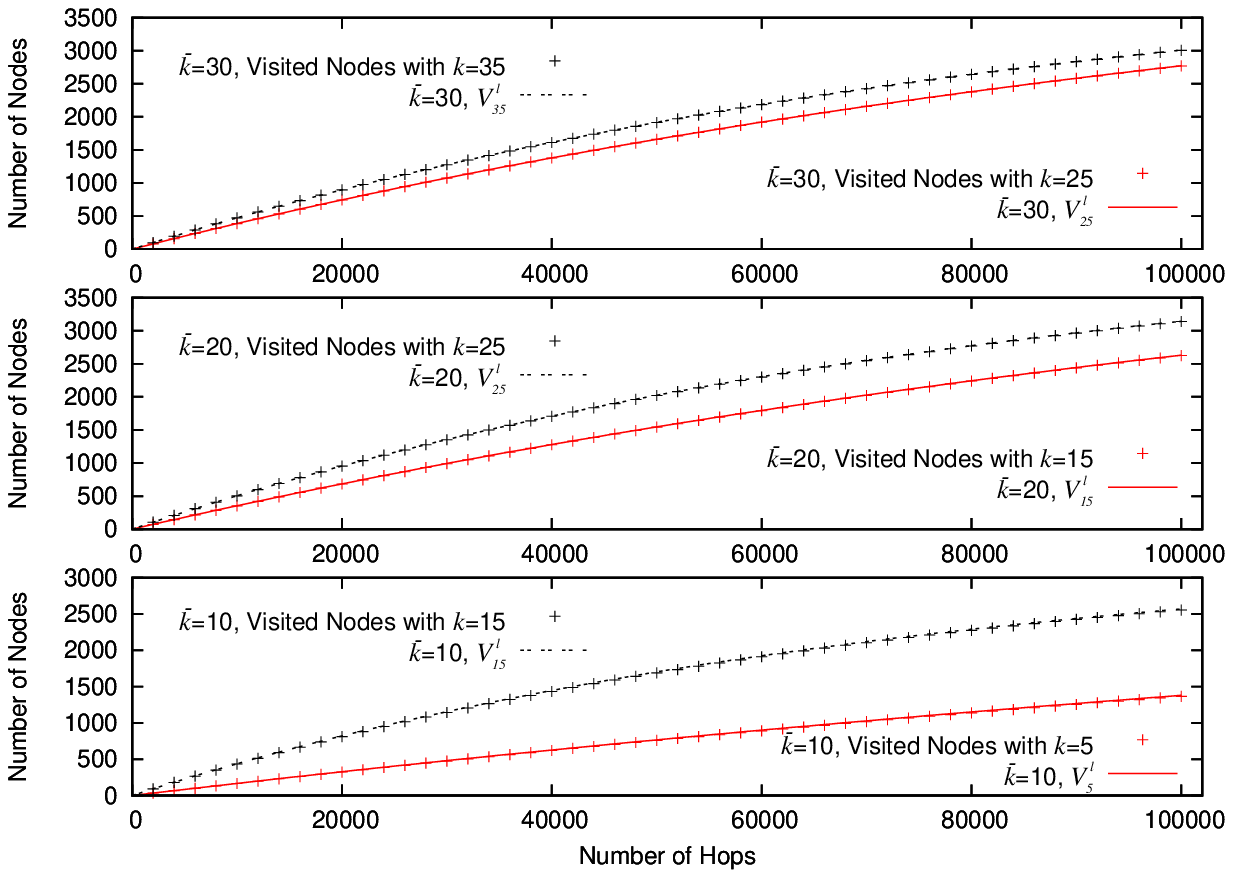}}
\label{fig:visitedPerDegRandom100}
}
\subfigure[Small-world; $n=10^5$; $\overline{k}=10$, $20$, $30$.]{
\resizebox{0.47\textwidth}{!}{\includegraphics{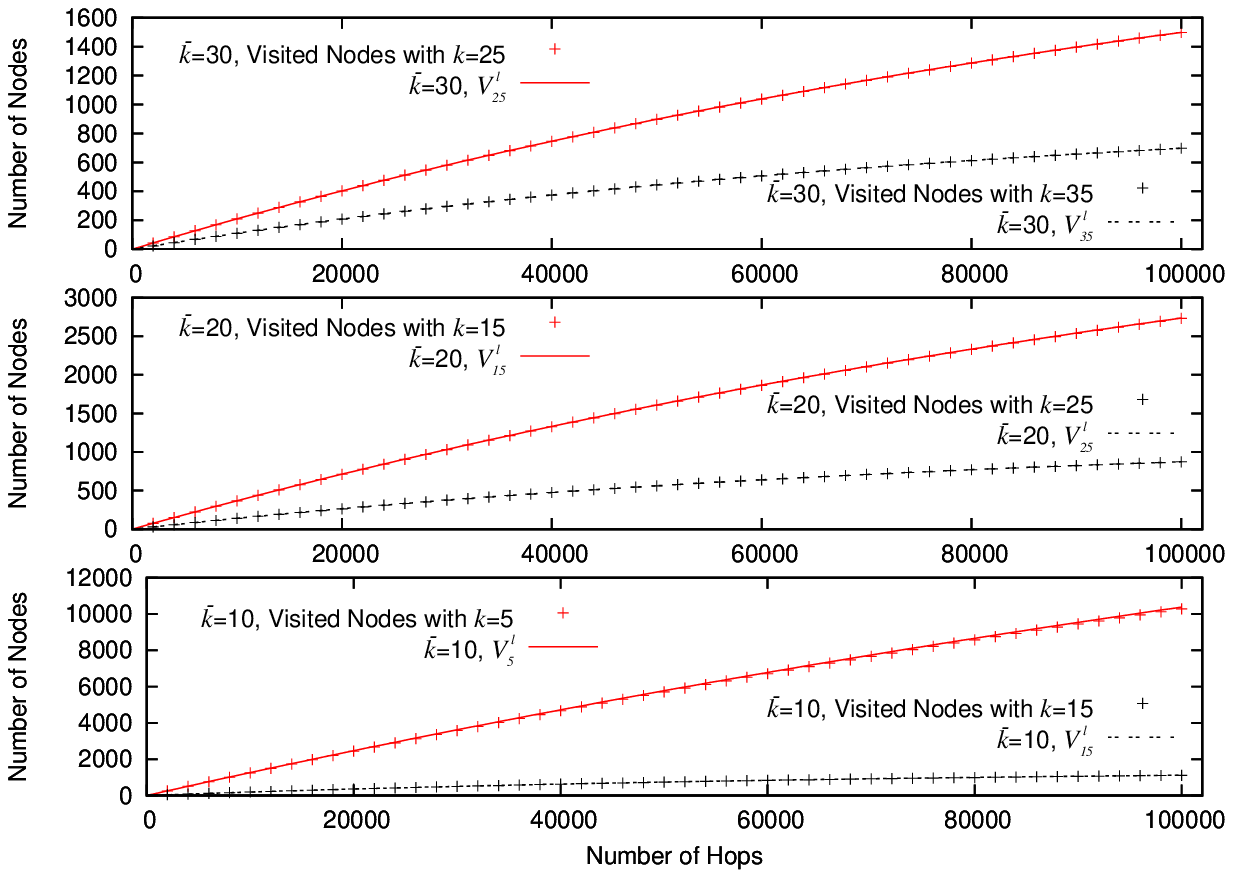}}
\label{fig:visitedPerDegSmallWorld100}
}
\caption{Visited nodes $V^l_k$, for $k=\overline{k}+5$
and $k=\overline{k}-5$.}
\label{fig:visitedPerDeg}
\end{center} 
\end{figure*}

\begin{figure*}[t]
\begin{center}
\subfigure[Erdos-Renyi and Small-world; $n=10^5$; $\overline{k}=10$, $30$.]{
\resizebox{0.47\textwidth}{!}{\includegraphics{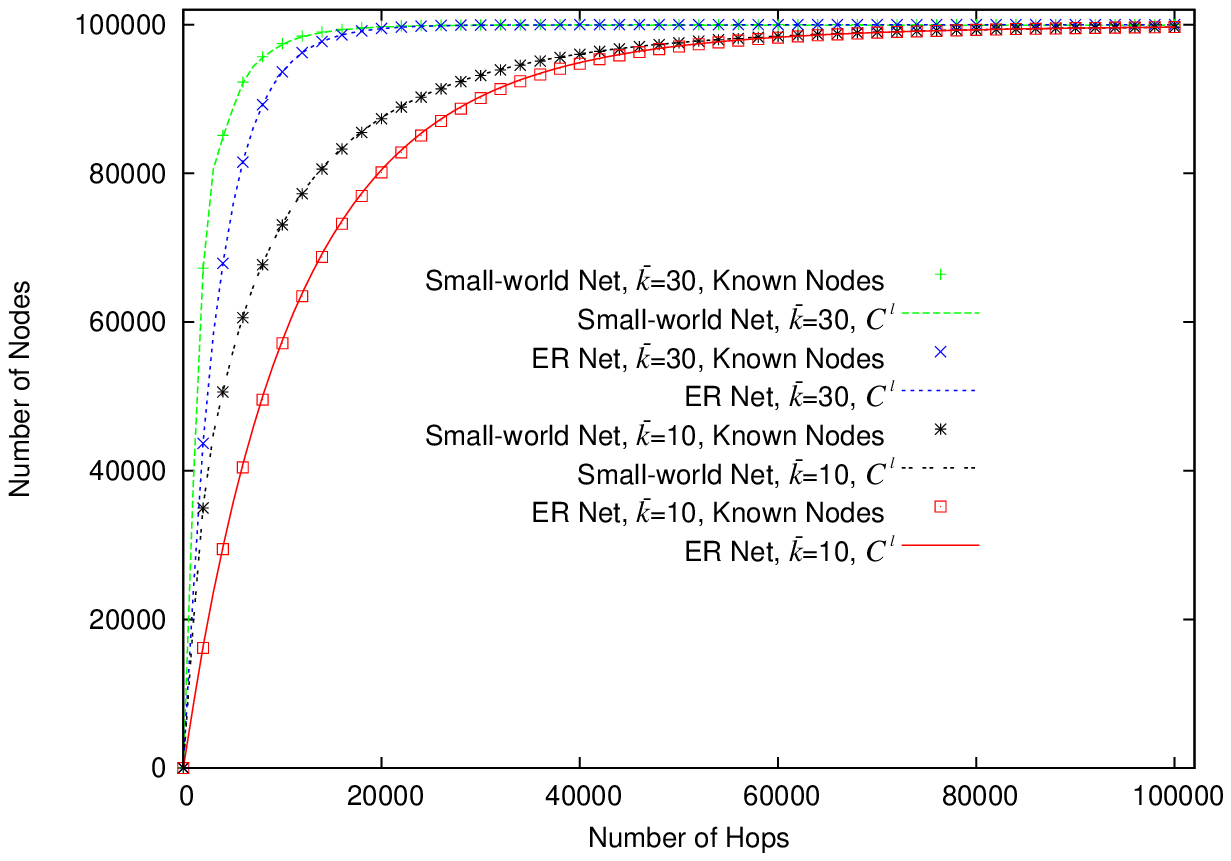}}
\label{fig:knownRamdomSmallWorld100}
}
\subfigure[Erdos-Renyi; $n=5\cdot 10^4$, $10^5$; $\overline{k}=10$, $30$.]{
\resizebox{0.47\textwidth}{!}{\includegraphics{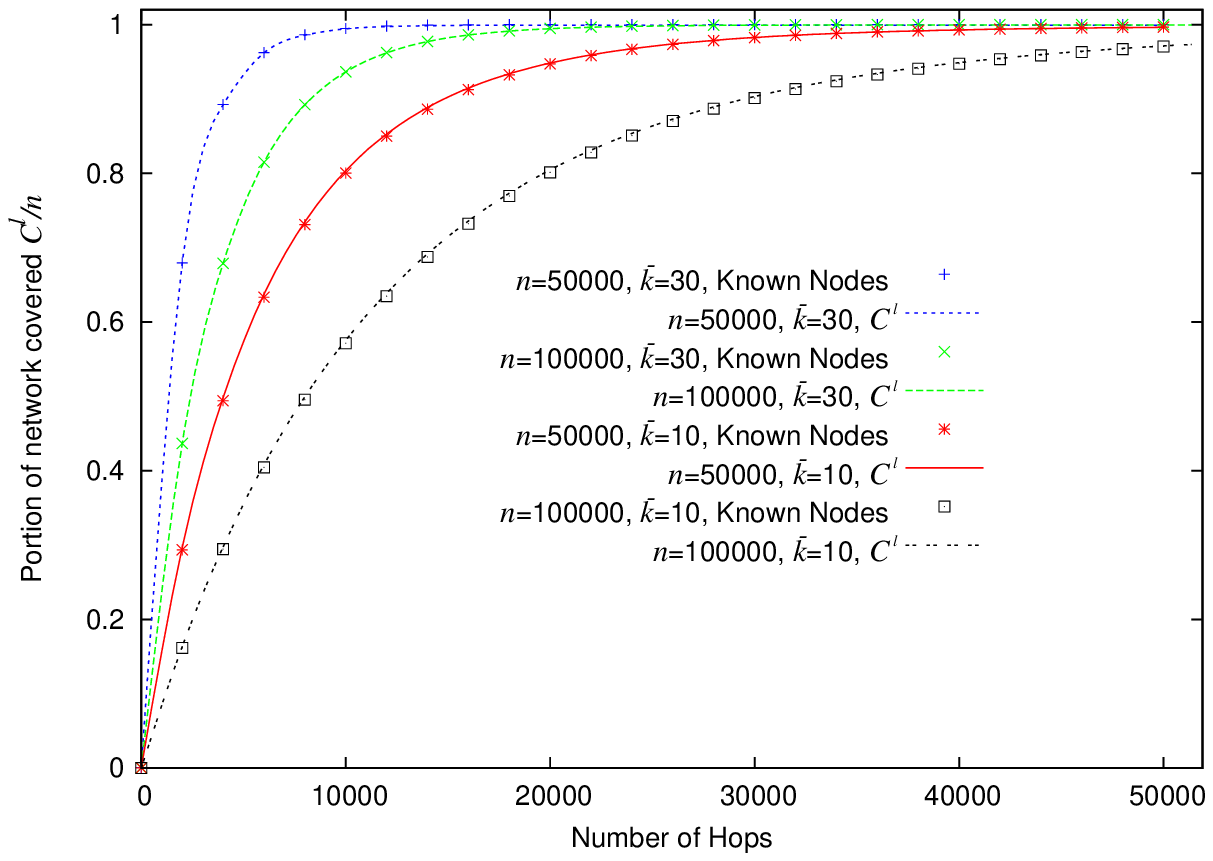}}
\label{fig:knownRandom50100}
}
\caption{Covered nodes $C^l$.}
\label{fig:known}
\end{center} 
\end{figure*}

\begin{figure*}[t]
\begin{center}
\subfigure[Erdos-Renyi; $n=10^5$; $\overline{k}=10$, $20$, $30$.]{
\resizebox{0.47\textwidth}{!}{\includegraphics{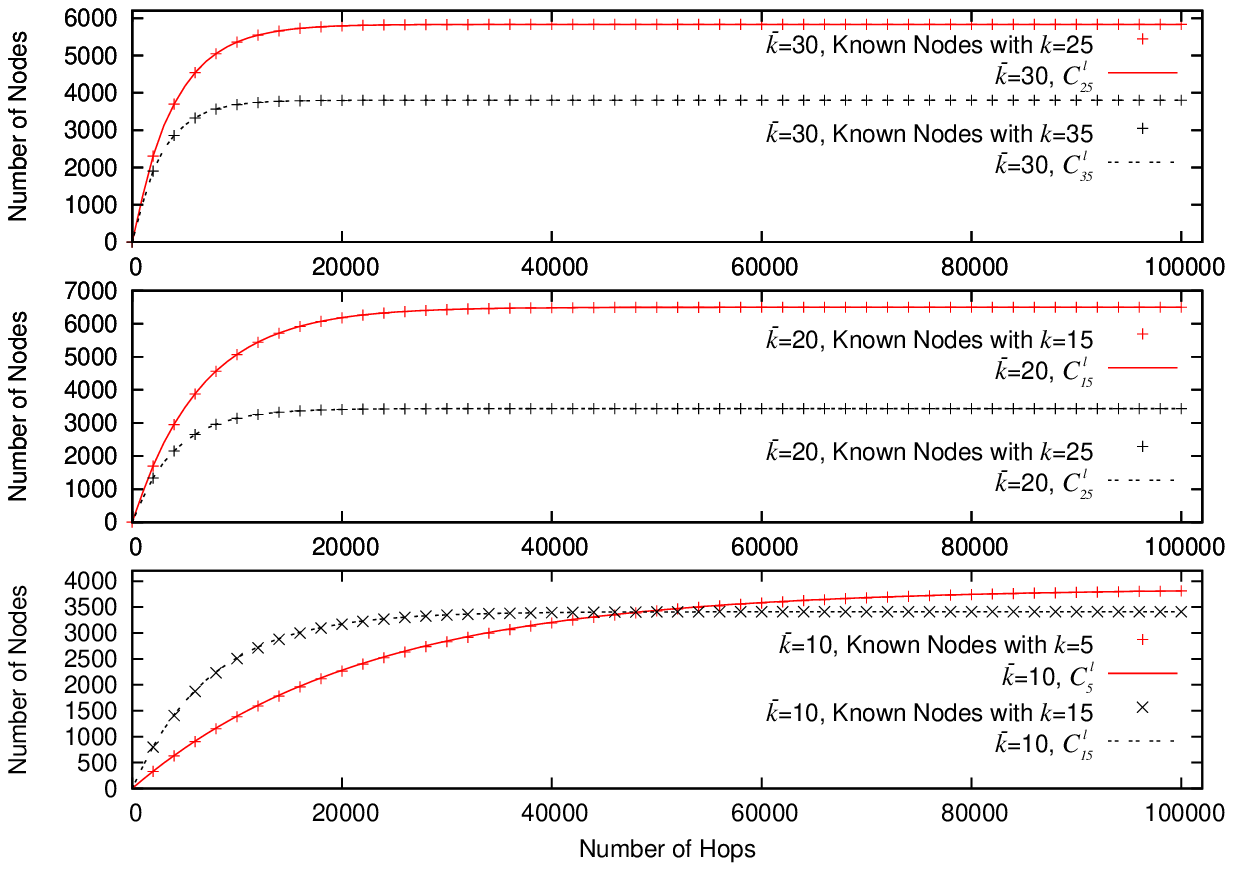}}
\label{fig:knownPerDegRamdom100}
}
\subfigure[Small-world; $n=10^5$; $\overline{k}=10$, $20$, $30$.]{
\resizebox{0.47\textwidth}{!}{\includegraphics{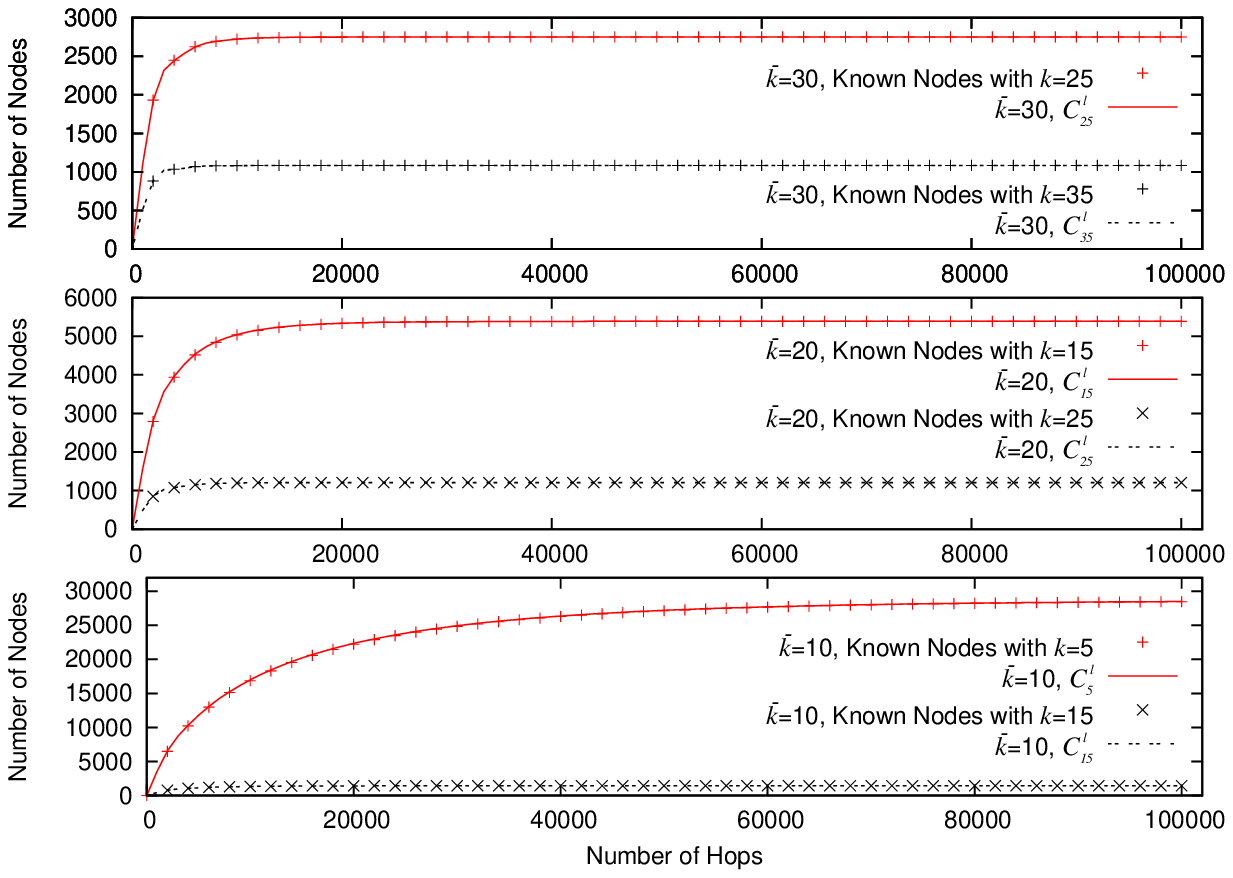}}
\label{fig:knowPerDegSmallWorld100}
}
\caption{Covered nodes $C^l_k$, for $k=\overline{k}+5$
and $k=\overline{k}-5$.}
\label{fig:knownPerDeg}
\end{center} 
\end{figure*}


\paragraph*{Number of Visited and Covered Nodes}

Our first goal is to study the evolution of the network coverage by random walks
in real networks. 

The experiments were run on networks of two sizes, $n=5\cdot 10^4$
and~$n=10^5$~nodes.  Networks were built using three different average
degrees:~$\overline{k}=10$, $\overline{k}=20$ and~$\overline{k}=30$. In each
network we ran $10^4$~random walks of length $n=|V|$. The source node of each
random walk was chosen uniformly at random. From the experiments, we obtained
the average number of visited and covered nodes for each degree $k$ at each hop
$l$.  Finally, for each network, we extracted its degree distribution $n_k$ and
apply the expressions described in the previous section to get a prediction of
those values, given by $V^l_k$ and $C^l_k$. Results are shown in
Figures~\ref{fig:visited},~\ref{fig:visitedPerDeg},~\ref{fig:known},
and~\ref{fig:knownPerDeg}. For the sake of clarity, the experimental results are
shown every 2000 hops in all figures. Model predictions, on the other hand, are
drawn as lines.

Figure~\ref{fig:visitedRandomSmallWorld50} shows the evolution of the number of
visited nodes in ER and small-world networks of size $n=5\cdot10^4$ nodes, with
two different average degrees $\overline{k}=10$ and $\overline{k}=30$. We see
that, although the length of the random walks is enough to potentially include
all the nodes, only a fraction of them are visited. This happens because of the
revisiting effect, and it is more evident when the number of hops increases,
since the probability of revisiting grows with the number of hops.  The
revisiting effect is stronger in small-world networks than in random networks.
The reason is the uneven distribution of the nodes degrees: there are some nodes
with a very high degree that will be visited once and again by the random walk.
Thus, the chances of finding new nodes at each hop are lowered faster in
small-world networks than in ER networks.  Also, we observe in
Figure~\ref{fig:visitedRandomSmallWorld50} that in networks of smaller
$\overline{k}$ the revisiting effect is stronger.  Finally,
Figure~\ref{fig:visitedRandomSmallWorld50100} shows the impact of the network
size $n$ on the amount of visited nodes. As expected, a greater $n$ implies a
lesser number of revisits for the same number of hops. In all cases, the
prediction $V^l$ of the total amount of different nodes visited is very close to
the experimental results.

In Figure~\ref{fig:visitedPerDeg} we study the accuracy of the predictions of
the amount of visited nodes of a particular degree $k$ at each hop $l$, $V^l_k$.
We draw the results and predictions of degrees $k=\overline{k}+5$ and
$k=\overline{k}-5$, for $\overline{k}=10$, $\overline{k}=20$ and
$\overline{k}=30$. Again, it can be seen that the model predictions fit very
well with the experimental results, despite the revisits and the different
behavior observed for different degrees.

Figure~\ref{fig:known} gives the results of the experiments run to study the
coverage of the random walk. Figure~\ref{fig:knownRamdomSmallWorld100} shows how
the coverage grows faster in small-world networks than in ER networks for
networks of the same average degree $\overline{k}$. This contrasts with the
amount of visited nodes, that behave in the opposite way (see previous
paragraphs). The reason is the presence of well-connected nodes, that are
quickly visited during the first hops of  the random walk and increase
considerably the coverage because of the high amount of neighbors they have. For
example, after $4000$ hops, the random walk has covered about half of the
small-world network with $\overline{k}=10$, while in the ER network of the same
$\overline{k}$ the random walk only has covered close to $30\%$ of the nodes.
Moreover, we can see that the network average degree has also an important
impact on the coverage. In both kind of networks the coverage grows faster when
the average degree is higher. Besides, we observe that the difference of the
coverage for both networks decreases more quickly for a higher $\overline{k}$.
Figure~\ref{fig:knownRamdomSmallWorld100} confirms the importance of the average
degree, comparing the results for networks of different size and $\overline{k}$.
In addition, Figure~\ref{fig:knownRandom50100} compares the results of the
coverage for ER networks of different sizes and average degrees.  As it could be
expected, the networks of smaller size require less hops to be covered. We
observe also that the average degree has an important influence on the coverage
difference.  The greater the average degree, the faster the coverage of both
networks converges. In all cases, the $C^l$ values given by the model predict
very well how the coverage behaves and evolves.

Figure~\ref{fig:knownPerDeg} allows to check the precision of the coverage
predictions for different $k$ values, $C^l_k$. As before, the values provided
are very close to the experimental results, although the behavior of the
coverage changes strongly depending on the kind of network and average degree.

\begin{figure*}[t]
\begin{center}
\subfigure[Erdos-Renyi and Small-world; $n=10^5$; $\overline{k}=10$.]{
\resizebox{0.47\textwidth}{!}{\includegraphics{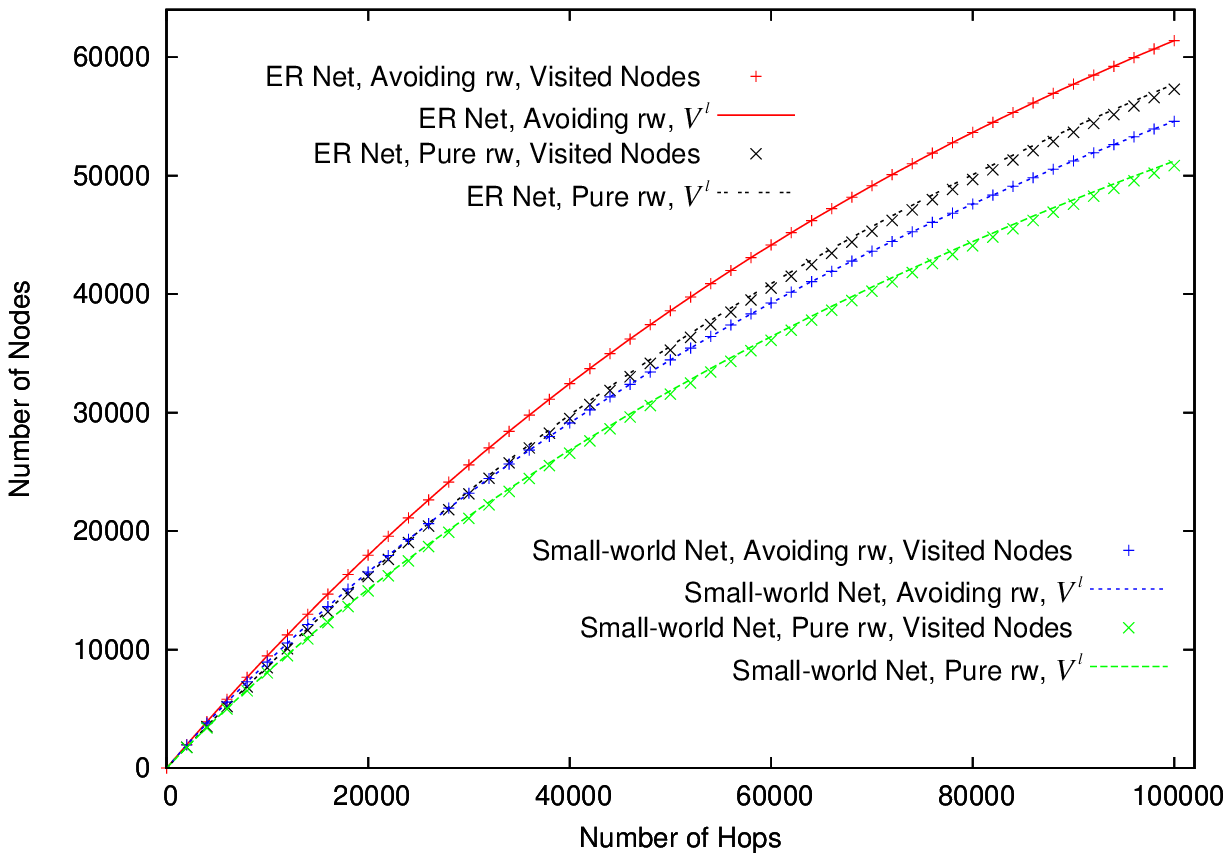}}
\label{fig:avoidVisited}
}
\subfigure[Erdos-Renyi and Small-world; $n=10^5$; $\overline{k}=10$.]{
\resizebox{0.47\textwidth}{!}{\includegraphics{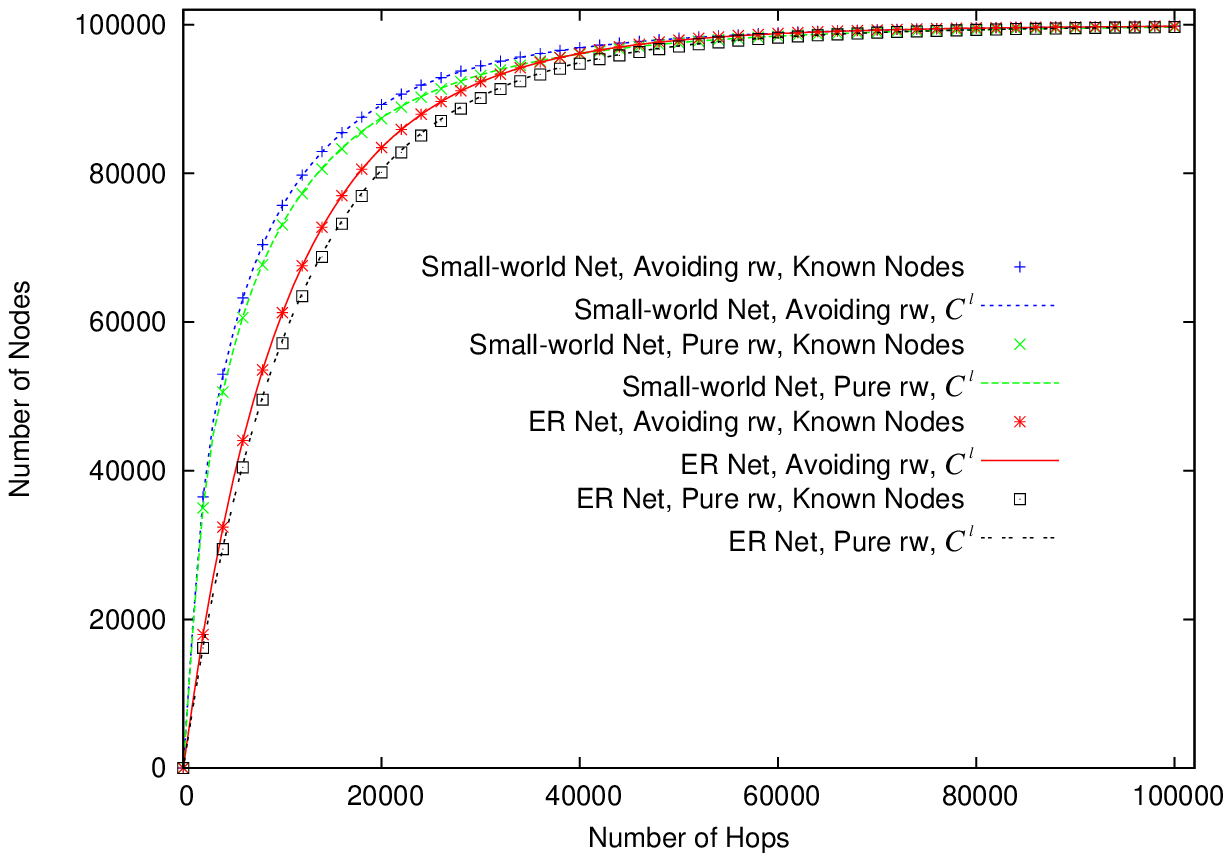}}
\label{fig:avoidKnown}
}
\caption{Avoiding Random Walk, Visited and Covered Nodes $V^l$ and $C^l$.}
\label{fig:avoid}
\end{center} 
\end{figure*}

Finally, we check the model accuracy for random walks that avoid the previous
node, the \emph{avoiding random walk}. As stated in
Section~\ref{subsec:visitedNodes}, the avoiding random walk can be easily
implemented by our model just by setting $P_b=0$ (see
Equation~\ref{equ:probBack}). Results are shown in Figure~\ref{fig:avoid}. There
we compare the coverage of pure and avoiding random walks in ER and small-world
networks of size $n=10^5$ nodes and average degree $\overline{k}=10$.
Figure~\ref{fig:avoidVisited} confirms that, as expected, the avoiding random
walk is able to visit a greater number of different nodes, as the revisiting
effect is, to a certain degree, lessened. However, Figure~\ref{fig:avoidKnown}
shows that this has little impact on the network coverage. We find that there is
only a small increase on the amount of covered nodes when using avoiding random
walks, for both kind of networks. Nonetheless, in all cases the $V^l$ and $C^l$
values given by the model are very close to real results.

\begin{figure*}[t]
\begin{center}
\subfigure[Erdos-Renyi; $n=10^4,\:...,2\cdot 10^5$; $\overline{k}=10,20,30$.]{
\resizebox{0.47\textwidth}{!}{\includegraphics{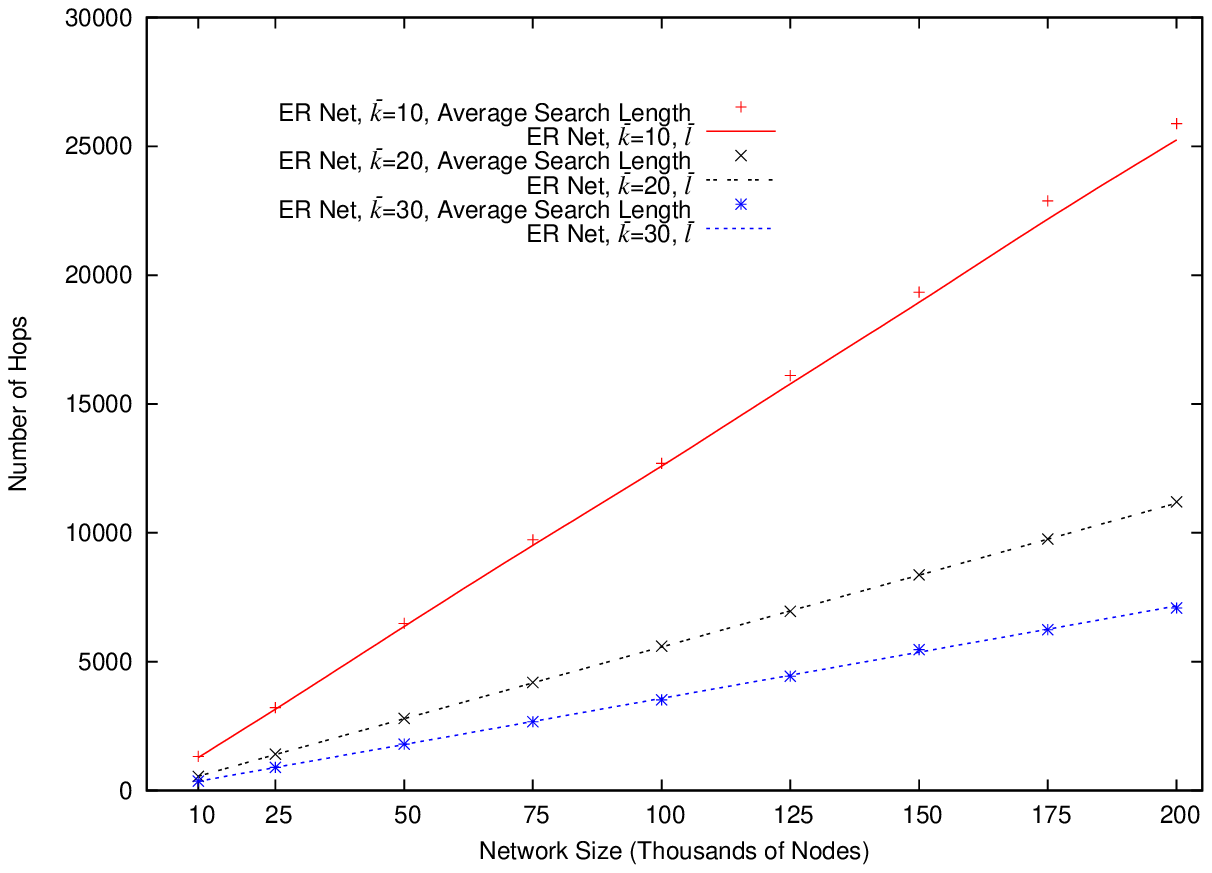}}
\label{fig:averSearchLenER}
}
\subfigure[Small-world; $n=10^4,\:...,2\cdot 10^5$; $\overline{k}=10,20,30$.]{
\resizebox{0.47\textwidth}{!}{\includegraphics{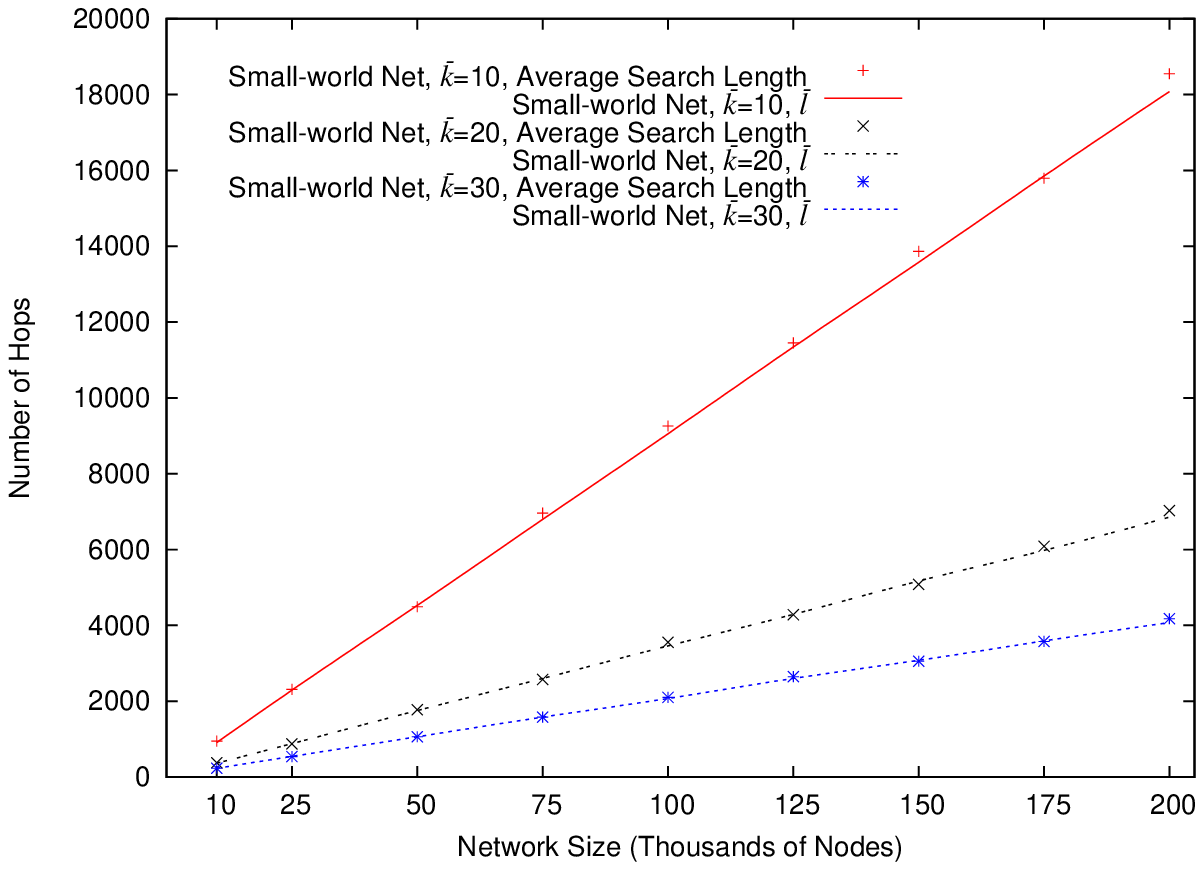}}
\label{fig:averSearchLenNewman}
}
\caption{Average Search Length $\overline{l}$.}
\label{fig:AverageSearchLength}
\end{center} 
\end{figure*}


\paragraph*{Average Search Length}
For the experiments regarding the average search length we used networks whose
sizes ranged from $10^4$ to $2\cdot 10^5$~nodes. In each experiment we ran
$10^4$~searches, averaging the obtained results. At each search, two nodes (one
corresponding to the source and the other to the destination) were chosen
uniformly at random. Starting from the source, a random walk traversed the
network until the destination node was found (i.e., a neighbor of the
destination is visited).

The first thing to note is that the average search length grows linearly with
the network size in both ER and small-world networks. Besides, the average
degree $\overline{k}$ has an important effect on the results. The bigger the
$\overline{k}$, the shortest the searches are. The reason is that a higher
$\overline{k}$ implies that at each hop more nodes of the network are
discovered. Also, it can be observed in Figure~\ref{fig:AverageSearchLength}
that the average search length is greater in ER networks than in small-world
networks. This can be explained if we take into account that random walks, on
average, cover more nodes in small-world networks than in ER networks (see
Figures~\ref{fig:known}).

As in the previous experiments, Figure~\ref{fig:AverageSearchLength} also shows
that our experimental results regarding the average search length correspond
very close to the analytical results that were obtained.

At this point, we would like to note that, given the assumptions we made in our
analytical model, it seems that the very good match achieved with the
experimental results could only occur if these assumptions are correct. As a
matter of fact, we have verified, in practice (see Figs.~\ref{fig:property}
and~\ref{fig:property2}), that the type of networks we consider in this paper,
indeed, fulfill our assumptions.

\begin{figure*}[t]
\begin{center}
\subfigure[Barabasi; $n=10^4,\:...,2\cdot 10^5$; $\overline{k}=10,20,30$.]{
\resizebox{0.47\textwidth}{!}{\includegraphics{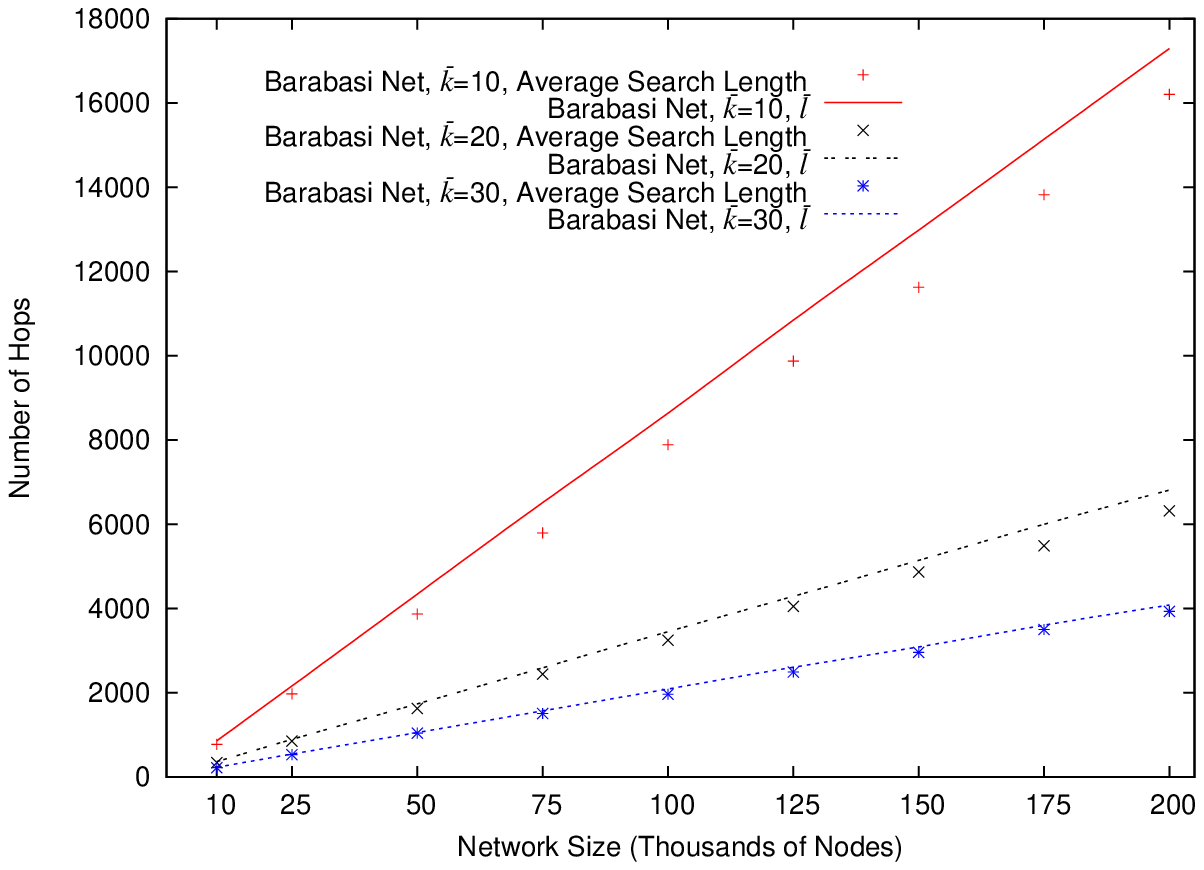}}
\label{fig:averSearchLenBar}
}
\subfigure[Toroid; $n=10^4,\:...,2\cdot 10^5$; $\overline{k}=10,16$.]{
\resizebox{0.47\textwidth}{!}{\includegraphics{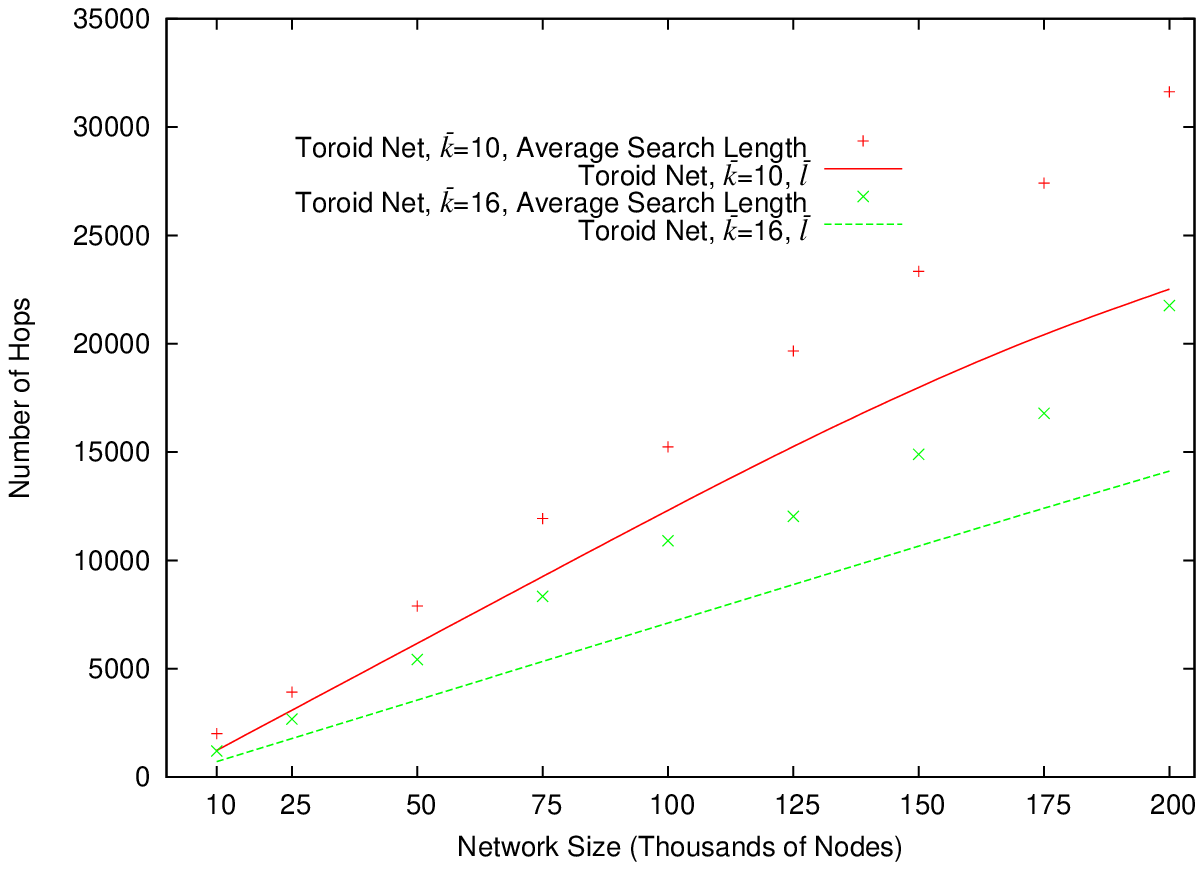}}
\label{fig:averSearchLenTor}
}
\caption{Average Search Length $\overline{l}$, not pure random networks.}
\label{fig:AverageSearchLengthNotRandom}
\end{center} 
\end{figure*}

On the other hand, it is clear that if we take into account networks that do not
fulfill some of our assumptions, then a certain mismatch should be expected.
For instance, networks built by preferential mechanisms are known not to
preserve the independence of degrees of neighbors~\cite{Krapivsky01}.
Therefore, we should not aim for a very close correspondence between analytical
and experimental results. We have performed the same experiments we ran for
random and small-world networks regarding the average search length, but this
time with networks built using the preferential attachment mechanism proposed by
Barab\'{a}si~\cite{Barabasi99}. Now, we have observed that, as expected, in
preferential networks our experimental results do not correspond very close to
the analytical results (see Fig.~\ref{fig:averSearchLenBar}).  Instead, the
model seems to be consistently pessimistic. Also, the error continuously grows
with the network size.

Finally, we have tested the model against Toroidal networks of different average
degrees $\overline{k}=10$ (5 dimensions) and $\overline{k}=16$ (8 dimensions).
Our intention is to analyze networks which are not random at all.  Results,
which are shown in Fig.~\ref{fig:averSearchLenTor}, show a very clear mismatch
among the results predicted by the model and the actual performance of the
random walk.

\section{Duration of Searches by Random Walks}
\label{sec:searchesByRW}

In this section, we present the second part of our model. Here we provide useful
expressions that allow to predict the performance of random walks as a search
tool, which is the main goal of this work. These expressions rely on the same
estimation of the average search length (like the one described in the previous
section), that is combined with Queuing Theory~\cite{Stallings00}. As a result,
given the processing capacities and degrees of nodes, we are able to compute two
key values:

\begin{itemize}
    \item The \emph{load limit}: the searches rate limit that the network can
    handle before saturation.
    \item The \emph{average search time}: the average time it takes to complete
    a search, given the global load.
\end{itemize}

Also, we show how these expressions can be used to analyze which features a
network should have so random walks have a better performance (i.e., searches
are solved in less time). In particular, we focus on studying the relationship
between degree and capacity distributions, showing that the minimum search time
is obtained when nodes of higher capacities are also those of higher degrees.

In our analysis, networks are assumed to be \emph{Jackson
networks}~\cite{Stallings00}: the arrival of new searches into the network
follows a Poisson distribution and the service at each node is a Poisson
process.


\subsection{Searches Length and Load on Nodes}
Our first step is to set the relationship between the average searches length
and the system load. Each search is processed, on average, $1 + \overline{l}$
times (once at the source node, and once at each step of the random walk). Using
this, we can express the total load on all the nodes of the system, $\lambda$,
as
\begin{equation}
    \label{equ:totalLoad}
    \lambda\:=\:(1+\overline{l})\:\gamma,
\end{equation}

where $\gamma$ is the load injected in the system by new searches, that we
assume to be known. Note that $\lambda$ is composed of the new generated
searches ($\gamma$), plus the searches that move from one node to another,
denoted by $\gamma'$. Hence,
\begin{equation}
\label{equ:lByLoads}
    \overline{l}\:=\:\frac{\gamma' + \gamma}{\gamma} - 1
    \:=\:\frac{\gamma'}{\gamma}.
\end{equation}

To compute the load on each particular node $j$, $\lambda_j$, let us take into
account that the probability that a random walk visits a node is proportional to
the node's degree (see Section~\ref{sec:knownNodesAverLeng}).  This implies
that, for each node $j \in V$, the load on node $j$ due to search messages,
denoted $\gamma_j'$, is proportional to its degree $k_j$. As a result, we have
that there is a value $\tau$ such that $\gamma_j'=\tau \; k_j$, for all $j$.
Hence, $\gamma'=\sum_j\gamma_j'=\tau \; d$, where $d$ is the sum of all degrees
in the network (i.e., $d=\sum_k n_k \; k$). Therefore,

\begin{equation}
\label{equ:lByLoads2}
    \tau\:=\:\frac{\overline{l} \, \gamma}{d}.
\end{equation}

Assuming that all nodes generate approximately the same number of new searches
($\gamma/n$), we can compute the average load at node~$j$ as
\begin{equation}
\label{equ:loadOnNode}
    \lambda_j\:=\:\tau \; k_j+\frac{\gamma}{n}
    \:=\: \gamma \left( \frac{\overline{l} \; k_j}{d}+\frac{1}{n} \right).
\end{equation}
where the first term represents the load  due to search messages, and the second
term to the searches generated at node~$j$. Note that any other search
generation rate model can be implemented just by changing the term $\gamma/n$.

\subsection{Average Search Duration}

In order to obtain the average search duration, $T_r$, we use \emph{Little's
Law}~\cite{Stallings00}, which states that
\begin{equation}
\label{equ:littleLaw2}
    r\:=\:\gamma\:\times\:T_r,
\end{equation}
where $r$ is the average number of \emph{resident} searches in the network
(i.e., searches that are waiting or being served), and $\gamma$ is the average
number of searches \emph{generated} per unit of time (i.e., the arrival rate of
searches). Observe that $\gamma$ is assumed to be known. Hence, the challenge to
compute $T_r$ is to obtain $r$. Let $r_j$ be the number of resident searches in
node~$j$. Then, $r\:=\:\sum_j {r_j}$.

To obtain $r_j$, we apply \emph{Little's Law} again, this time individually to
each node $j$:
\begin{equation}
\label{equ:littleLaw3}
    r_j\:=\:\lambda_j\:\times\:T_{r}^j,
\end{equation}
where $T_{r}^j$ is the average search time at node $j$ and $\lambda_j$ is the
average  \emph{load} at node~$j$, which includes both searches generated at
node~$j$ and searches due to messages from other nodes. Next we use that, by
\emph{Jackson's Theorem}~\cite{Giambene05} (recall we assume the network to be a
Jackson network), each node $j$ can be analyzed as a single M/M/1 queue with
Poisson arrival rate $\lambda_j$ and exponentially distributed service time with
mean $T_s^j$ (which can be computed from the node capacity, that we assume to be
known). Then:
\begin{equation}
\label{equ:resTimeNode}
    T_{r}^j\:=\:\frac{T_{s}^j}{1-\rho_j},
\end{equation}
where $\rho_j$ is the utilization rate and $T_{s}^j$ is the average service time
at node~$j$. As $\rho_j=\lambda_j \; T_{s}^j$, we can write
\begin{equation}
\label{equ:resTimeNode2}
    T_{r}^j\:=\:\frac{T_{s}^j}{1-\lambda_j \; T_{s}^j}.
\end{equation}

Once we have $\lambda_j$ and $T_r^j$, we can combine them to obtain
\begin{equation}
\label{equ:resTimeSystem}
\begin{split}
    T_r & =  \frac{r}{\gamma} \\
     & = \frac{\sum_j {r_j}}{\gamma}  \\
     & =  \frac{1}{\gamma}\sum_j{\lambda_j \; T_{r}^j} \\
     & = \frac{1}{\gamma}\sum_j{\frac{\lambda_j \; T_{s}^j}{1-\lambda_j \;
     T_{s}^j}} \\
     & = \sum_j{ \frac{T_{s}^j\left(\frac{\overline{l} \;
     k_j}{d}+\frac{1}{n}\right)}{1-T_{s}^j \; \gamma\left(\frac{\overline{l} \;
     k_j}{d}+\frac{1}{n}\right)} } \\
     & = \sum_j{\left( \frac{n \; d}{T_{s}^j\left( k_j \; \overline{l} \; n  + d
     \right)}-\gamma \right)^{-1}}.
\end{split}
\end{equation}

That is, we have provided an expression that computes the \emph{average search
time} using the topology, the average service times of nodes, and the search
arrival rate.


\subsection{Load Limit}
\label{sec:loadlimit}

Implicitly, in our previous results it has been assumed that no node is
overloaded (i.e.,  $\lambda_j < 1/T_s^j$ for all~$j$). Otherwise, the network
would never reach a stable state. Thus, a key value for any network is its
\emph{load limit}: the minimum search arrival rate ($\gamma$) that would
overload the network, denoted by $\gamma_o$.  Clearly, $\gamma_o =
\min_{j}\{\gamma_o^j\}$ being $\gamma_o^j$ the minimum search arrival rate that
would overload node~$j$.

From Equation~\ref{equ:loadOnNode}, we have that
\begin{equation}
\label{eq:min1}
    \lambda_j \:=\:k_j \; \frac{\overline{l}\; \gamma}{d}+\frac{\gamma}{n}.
\end{equation}

Also, since no node must be overloaded, it must be satisfied that
\begin{equation}
\label{eq:min2}
    \lambda_j \:<\:\frac{1}{T_s^j}.
\end{equation}

Combining Equation~\ref{eq:min1} with Equation~\ref{eq:min2} we have that, for
each~$j$, the following must hold:
\begin{equation}
    \gamma\:<\:\frac{d \; n}{T_s^j \; (k_j \; \overline{l} \; n + d)}.
\end{equation}

Therefore, the load limit for node~$j$ is
\begin{equation}
    \gamma_o^j \:=\:\frac{d \; n}{T_s^j \; (k_j \; \overline{l} \; n + d)},
\end{equation}

and
\begin{equation}
    \gamma_o \:=\: \min_{j} \left\{ \frac{d \; n}{T_s^j \; (k_j \; \overline{l}
    \; n + d)} \right\}.
\end{equation}


\subsection{Experimental Evaluation}

\paragraph*{Average Search Duration}
\begin{table}
\renewcommand{\arraystretch}{1.3}
\caption{Capacity distributions}
\label{table:timeModelCapacities}
\begin{center}
\begin{tabular}{|p{2.1cm}||p{2.1cm}|}
\hline
Percentage \newline \centering of nodes & Processing \newline  capacity  \\
\hline
 \centering 20\% & 1 \\
 \centering 45\% & 10 \\
 \centering 30\% & 100 \\
 \centering 4.9\% & 1,000 \\
 \centering 0.1\% & 10,000 \\
\hline
\end{tabular}
\end{center}
\end{table}

\begin{figure}[t]
   \centering
    \includegraphics[width=9cm]{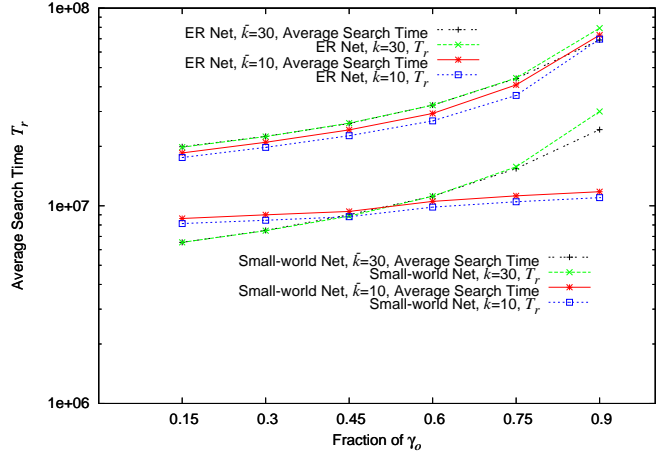}
    \caption{Average Search Times. For the analytical values ($T_r$), we
    used Equation~\ref{equ:resTimeSystem}, taking into account that $T_s^j$
    follows an exponential distribution with average $\lambda_j$ (i.e., $T_s^j
    \:\sim\:\text{Exponential}(\lambda_j)$), where $\lambda_j$ can be computed
    as the relation between the number of resources known and their processing
    capacity.}

\label{fig:timeModel}
\end{figure}

In this subsection, we present the results of a set of experiments addressed to
evaluate, in practice, the accuracy of our model for the average search time. As
in the previous experiments (Section~\ref{sec:experimentalResults}), we
conducted extensive simulations over ER and small-world networks. All networks
are made up of $10^4$~nodes.

In each experiment, nodes generate new searches following a Poisson process with
rate $\gamma/n$, where $\gamma$ is the global load on the network. When a node
starts a search for a resource, it first checks whether it already knows that
resource (i.e., if the node itself or any of its neighbors hold the resource).
If so, the search ends successfully. Otherwise, a search message for the
requested resource is created and sent to some neighbor node chosen uniformly at
random. When a node receives a search message, it also verifies whether it knows
the resource.  If so, the search is finished. Otherwise, the search is again
forwarded to another neighbor chosen uniformly at random. The experimental
results are obtained by averaging the results that were obtained.

We used six different global loads ($\gamma$): $0.15\times\gamma_o$,
$0.3\times\gamma_o$, $0.45\times\gamma_o$, $0.6\times\gamma_o$,
$0.75\times\gamma_o$ and $0.9\times\gamma_o$, where $\gamma_o$ is the minimum
arrival rate that would overload the network (see Section~\ref{sec:loadlimit}).
The distribution of the nodes search processing capacities $c_i$ is derived from
the measured bandwidth distributions of Gnutella~\cite{Saroiu02} (see
Table~\ref{table:timeModelCapacities}). Capacities are assigned so that nodes
with a higher degree are given a higher capacity. All nodes are assumed to have
the same number of resources $w=10,000$.  Each resource is held by one node, and
all resources have the same probability of being chosen for search.  The
processing time at each node $i$ follows an exponential distribution with an
average service time computed as $T_s^i=w \; k_i/c_i$. This average is computed
dividing the amount of resources checked for each search (the total amount of
resources known, $w(k+1)$, minus the resources of the node the search message
came from, $w$) by the node's capacity.

For each load, we measured the average search times experimentally for each
network.  Results are shown in Fig.~\ref{fig:timeModel}. It can be seen that, as
expected, the average search time always increases with the load, undergoing a
higher growth when it approaches the maximum arrival rate.  Furthermore, our
experimental results show a very close correspondence with the analytical
results that were obtained.

\paragraph*{Load Limit}
We have computed the $\gamma_o$ values for random and small-world networks with
different average degrees. For each kind of network and average degree five
networks were built with the capacity distribution presented in
Table~\ref{table:timeModelCapacities}. Our goal was to observe the variation of
the $\gamma_o$ for networks of the same type and $\overline{k}$, and also to
study the difference among the $\gamma_o$ values depending on the network kind
and average degree. 

\begin{figure}[t]
   \centering
    \includegraphics[width=9cm]{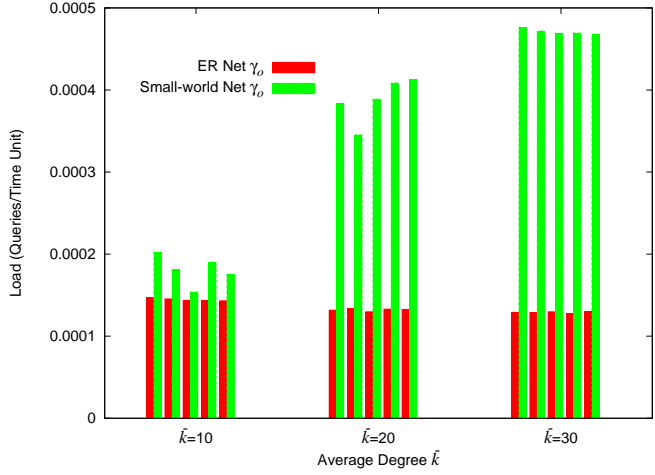}
    \caption{Load limit $\gamma_o$. Five different networks are created for each
    network type (ER or small-world) and average degree $\overline{k}$. The
    resulting $\gamma_o$ are shown grouped by $\overline{k}$.}
\label{fig:maxLoad}
\end{figure}

Results, which are shown in Figure~\ref{fig:maxLoad}, differ for random and
small-world networks. The first thing to note is that small-world networks can
handle a greater load than random networks.

Small-world networks present variations of the $\gamma_o$ values even for
networks of the same average degree. Despite this variation, it is clear that
the load limit tends to grow with the $\overline{k}$. The reason is that a
greater $\overline{k}$ implies a smaller global load for the same rate of
queries injected to the system. Recall that the total load is given by
$(1+\overline{l})\gamma$ (Equation~\ref{equ:totalLoad}) and that higher average
degrees lead to lesser average searches lengths $\overline{l}$
(Figure~\ref{fig:averSearchLenNewman}). Hence, it is possible to perform more
queries before overloading the network.

Erdos-Renyi networks however behave in a very different manner.  They present
very little variations of the $\gamma_o$ values. And, more surprising, there is
a small decrease of the load limit when the $\overline{k}$ grows. This contrasts
with the behavior of small-world networks. As it is shown in
Figure~\ref{fig:averSearchLenER}, larger average degrees imply smaller average
searches lengths and so a smaller global load. However, the $\gamma_o$ that can
be handled by the network does not change accordingly to this. The reason seems
to be that in ER networks the load is more evenly distributed among nodes.  This
implies that low capacity nodes have to handle an important amount of searches.
Besides, a greater average degree impacts on the average services times $T_s$ of
these nodes, as they know, and so they have to process, more resources per
search. Hence, these nodes keep being the bottleneck of the network despite the
smaller average search length, preventing the system to be able to handle a
greater load.

However, it is important to recall that these results are also due to the
capacity distribution used, and how it was distributed among the nodes. In
small-world networks, if we assign low capacities to high degree nodes we can
expect them to become bottlenecks of the network that force small $\gamma_o$
values. In ER networks, adding more high capacity nodes could change the
$\gamma_o$ tendency so it would grow with the average degree.  Exploring all
these phenomena is beyond the scope of this paper.

\subsection{Optimal Relationship between Degree and Capacity Distributions}
\label{subsec:relPerDegCap}

In this section we show that, when there is a full correlation between the
\emph{capacity} of a node (i.e., the number of searches a node can process per
time unit) and its degree, this leads to a minimal value of the average search
time $T_r$.

Let us first state the relation we assume between the capacity $c_j$ and the
average service time $T_s^j$ of a node $j$. We assume that the first is a
parameter that does not depend on the degree or the number of resources known by
the node, and only depends on the processor and network connection speeds. We
assume that the second is a strictly increasing function of the node's degree
$f(k_j)$. We assume that a node's service time is directly proportional to its
degree and inversely proportional to its capacity as follows:
\begin{equation}
\label{equ:serviceTime}
    T_{s}^j\:=\:\frac{f(k_j)}{c_j}.
\end{equation}

Let us now consider a pair of nodes $i,j \in V$, such that $k_j > k_i$ (so
$f(k_j)>f(k_i)$), and two possible positive capacities $c_1$ and $c_2$, such
that $c_1 > c_2$. We show that, if no other degree or capacity assignment
changes, having $c_j=c_1$ and $c_i=c_2$ gives a smaller average search time,
$T_r$, than the average search time $T'_r$ with reverse assignment $c'_j=c_2$
and $c'_i=c_1$.

Using Eq.~\ref{equ:serviceTime}, we obtain the following possible average
service times:
\begin{equation}
\label{equ:serviceTimes}
T_{s,1}^j=\frac{f(k_j)}{c_1}\:;\:T_{s,1}^i=\frac{f(k_i)}{c_2}\:;\:
T_{s,2}^j=\frac{f(k_j)}{c_2}\:;\:T_{s,2}^i=\frac{f(k_i)}{c_1},
\end{equation}
in which $T_{s,1}$ are the service times obtained with the first capacity
assignment and $T_{s,2}$ are the service times obtained with the second. From
the above equations, we have
\begin{equation}
\label{equ:serProdEqual}
\begin{split}
    T_{s,1}^jT_{s,1}^i\: & =\:\frac{f(k_i) \; f(k_j)}{c_1 \; c_2} \\
    & =\:T_{s,2}^j \; T_{s,2}^i,
\end{split}
\end{equation}
and
\begin{equation}
\label{equ:servTimes}
\begin{split}
    T_{s,1}^i-T_{s,2}^i\: & =\:f(k_i) \; \frac{c_1-c_2}{c_1 \; c_2} \\
    & <\: f(k_j) \; \frac{c_1-c_2}{c_1 \; c_2} \\
    & =\:T_{s,2}^j-T_{s,1}^j.
\end{split}
\end{equation}

Let $\lambda_i$ and $\lambda_j$ be the loads on $i$ and $j$. Since $k_i<k_j$,
then $\lambda_i<\lambda_j$. Hence, from this and Eq.~\ref{equ:servTimes}, we
find that
\begin{equation}
\label{equ:serTimesDif}
    \lambda_i \; T_{s,1}^i+\lambda_j \; T_{s,1}^j\:<\:\lambda_i \;
    T_{s,2}^i+\lambda_j \; T_{s,2}^j.
\end{equation}
To compute the values $T_r$ and $T'_r$, we use Eq.~\ref{equ:resTimeSystem}
\begin{equation}
    T_r\:=\:\frac{1}{\gamma}\left(r_i+r_j + \sum_{h\neq i,h\neq j}{r_h}\right);
\end{equation}
\begin{equation}
    T'_r\:=\:\frac{1}{\gamma}\left(r'_i+r'_j + \sum_{h\neq i,h\neq
    j}{r_h}\right),
\end{equation}
where $r_i$ and $r_j$ are obtained with the first capacity assignment and $r'_i$
and $r'_j$ with the second. Observe that $r_h$ remains the same for any node $h$
that is neither $i$ nor $j$, because its degree, load, and capacity are just the
same for both cases. Hence,  if
$r_i+r_j<r'_i+r'_j$ then $T_r<T'_r$. 

From Eqs.~\ref{equ:littleLaw3} and \ref{equ:resTimeNode2}, we obtain that
\begin{equation}
\begin{split}
r_i+r_j & = \frac{\lambda_i \; T_{s,1}^i}{1-\lambda_i \; T_{s,1}^i}+
            \frac{\lambda_j \; T_{s,1}^j}{1-\lambda_j \; T_{s,1}^j} \\
        & = \frac{-2 \; \lambda_i \; \lambda_j \; T_{s,1}^i \; T_{s,1}^j
                  +\lambda_i \; T_{s,1}^i+\lambda_j \; T_{s,1}^j}
                 {1+\lambda_i \; \lambda_j \; T_{s,1}^i \; T_{s,1}^j
                  -(\lambda_i \; T_{s,1}^i+\lambda_j \; T_{s,1}^j)},
\end{split}
\end{equation}
and
\begin{equation}
\begin{split}
r'_i+r'_j & = \frac{\lambda_i \; T_{s,2}^i}{1-\lambda_i \; T_{s,2}^i}+
              \frac{\lambda_j \; T_{s,2}^j}{1-\lambda_j \; T_{s,2}^j}\\
          & = \frac{-2 \; \lambda_i \; \lambda_j \; T_{s,2}^i \; T_{s,2}^j+
                    \lambda_i \; T_{s,2}^i+\lambda_j \; T_{s,2}^j}
                   {1+\lambda_i \; \lambda_j \; T_{s,2}^i \; T_{s,2}^j-
                    (\lambda_i \; T_{s,2}^i+\lambda_j \; T_{s,2}^j)}.
\end{split}
\end{equation}

Finally, applying Eqs.~\ref{equ:serProdEqual} and \ref{equ:serTimesDif}, we
conclude that
\begin{equation}
    r_i+r_j\:<\:r'_i+r'_j,
\end{equation}
and hence
\begin{equation}
    T_r\:<\:T'_r.
\end{equation}

This proves that, for a given degree distribution, the best performance will be
obtained by assigning the largest capacities to the nodes with the largest
degrees. Note that we have found a condition that is necessary in order to
attain the minimum possible $T_r$, once the degree distribution has been set.
However, different degree distributions can obtain very different $T_r$ values.


\section{Conclusions}
\label{sec:conclusions}

In this paper, we have presented an analytical model that allows us to predict
the behavior of random walks. Furthermore, we have also performed some
experiments that confirm the correctness of our expressions.

Some work can be carried out to complement our results. For instance, several
random walks can be used at the same time, a situation that could be used to
further improve the efficiency of the search mechanism. These random walks could
run independently or, in order to cover separated regions on the graphs,
coordinate among them in some way.

\bibliographystyle{elsart-num}

\end{document}